\documentclass[prb,aps,twocolumn,showpacs,floatfix]{revtex4}

\usepackage{amsmath,amssymb}
\usepackage{graphicx}
\usepackage{version}
\DeclareGraphicsExtensions{.eps,.ps}

\newcommand\rv{\vec{r}}
\newcommand\rprime{\rv^{\,\prime}}

\newcommand\deltaR{\delta R}

\newcommand\Vsc{\tilde{V}}
\newcommand\Vo[1]{V^{(0)}(#1)}
\newcommand\Vq[1]{\hat{V}(#1)}

\newcommand\eps[1]{\varepsilon_{#1}}

\newcommand\epso[1]{\epsilon^{(0)}(#1)}
\newcommand\epsq[1]{\hat{\epsilon}(#1)}

\begin{document}

\title
{Stability of Metal Nanowires at Ultrahigh Current Densities}
\author{C.-H.\ Zhang}
\affiliation{Department of Physics, University of Arizona,
1118 E.\ 4th Street, Tucson, AZ 85721}
\author{J.\ B\"urki}
\affiliation{Department of Physics, University of Arizona,
1118 E.\ 4th Street, Tucson, AZ 85721}
\author{C.\ A.\ Stafford}
\affiliation{Department of Physics, University of Arizona,
1118 E.\ 4th Street, Tucson, AZ 85721}

\date{Submitted: Nov. 4, 2004; Resubmitted: December 21, 2004;
Modified: \today}

\begin{abstract}
We develop a generalized grand canonical potential for the ballistic
nonequilibrium electron distribution in a metal nanowire with a finite applied
bias voltage.  Coulomb interactions are treated in the self-consistent Hartree
approximation, in order to ensure gauge invariance.  
Using this formalism, we investigate the stability and cohesive
properties of metallic nanocylinders at ultrahigh current densities.
A linear stability analysis shows that metal nanowires with certain {\em magic 
conductance values} can support current densities up to
$10^{11} \mbox{A}/\mbox{cm}^2$, which would vaporize a macroscopic 
piece of metal.  
This finding is consistent with experimental studies of gold nanowires.
Interestingly, our analysis also reveals the existence of reentrant stability
zones---geometries that are stable only under an applied bias.  
\end{abstract}


\maketitle

\section{Introduction}\label{sec:introduction}

Metal nanowires have been the subject of  many experimental and theoretical
studies, both for their unique properties and potential applications (see
Ref.\ \onlinecite{Agrait03} for a review of the field).
One of the most remarkable properties of metal nanowires is their ability to
support extremely high current densities without breaking apart or vaporizing.
\cite{Yasuda97,Itakura99,Untiedt00,Hansen00,Yuki01,Mehrez02,Agrait02}
For noble metals, experiments can be carried out in air, and the first
few peaks in conductance histograms can withstand applied
voltages as high as one volt, and even two volts for the first peak,
corresponding to one conductance quantum $G_0=2e^2/h$.
Let us estimate the corresponding current density:
For a ballistic metallic conductor in the form of a cylinder of radius $R$,
the electrical conductance $G$ is given approximately by the Sharvin formula
$G \simeq G_0 (k_F R/2)^2$, where $k_F$ is the Fermi wavevector.
Therefore the current density at applied voltage $V$ is
\begin{equation}
j = \frac{GV}{\pi R^2}
\simeq \frac{k_F^2 G_0 V}{4 \pi} 
= \frac{3n e v_F}{8} \times \frac{eV}{\varepsilon_F},
\label{eq:bigj}
\end{equation}
where $n$ is the number density of conduction electrons, $v_F$ is the Fermi
velocity, and $\varepsilon_F$ is the Fermi energy.
For an applied bias of a few Volts, the factor $eV/\varepsilon_F$ is of
order unity, and the current density is of order 
$10^{11}$A/cm$^2$.
Such high current densities would vaporize a macroscopic wire, thus prompting 
questions on the reason for the remarkable stability of metal nanowires.

The first part of the answer to this question is that metal nanowires are 
typically shorter than the mean-free path $L_{\rm in}$ for inelastic scattering,
so that the conduction electrons can 
propagate through the wire without generating excitations such as 
phonons\cite{Agrait02} that heat the wire.  Instead, most of the dissipation 
takes place in the macroscopic contacts for the outgoing electrons.  However,
the absence of equilibration of the electron distribution within the nanowire 
raises another, more fundamental, question:  
What is the effect of a highly nonequilibrium electron
distribution on the stability of a metal nanowire, given that the conduction
electrons play a dominant role in the cohesion of metals?
That is the question to which the present article is devoted.

Under a finite bias, the scattering states of right- and left-moving
electrons in a nanowire are populated differently, even if there
is no inelastic scattering within the wire. An adequate
treatment of the electron-electron interactions
is crucial to correctly describe this nonequilibrium electron
distribution.
Some studies of transport\cite{Pascual97,Bogachek97} and cohesion\cite{Zagoskin98} 
in metal nanowires at finite bias did not include electron-electron interactions, so that the calculated transport and energetics depended separately
on both the left and right chemical potentials $\mu_+$ and $\mu_-$,
thus violating the {\em gauge invariance} condition:
The calculated physical quantities should depend only on the
voltage difference $eV=\mu_+-\mu_-$, and should be invariant under
a global shift of the electrochemical potential, since the total charge is
conserved.\cite{Christen96}
A self-consistent formulation of transport and cohesion at finite
bias has recently been developed based on {\it ab initio} and tight-binding
methods.\ \cite{Todorov00,Ventra01,Brandbyge02,Mehrez02,Mozos02}
These computational
techniques are particularly well-suited to the study of atomic chains,
but can become intractable for larger nanostructures.
An analytical approach to this problem is needed 
to study the interesting mesoscopic effects\cite{Urban04a,Stafford97a}
which occur in systems intermediate
in size between the macroscopic and the atomic scale.

In this paper, we extend our continuum model
\cite{Stafford97a,Kassubek99,Stafford99,Kassubek01,Zhang03,Burki03,Urban04b,Stafford00,Stafford01}
of metal nanowires to treat the ballistic nonequilibrium electron distribution
at finite bias.  
Our model provides a generic description of nanostructures formed of simple,
monovalent metals.  It is especially suitable for alkali metals, but is also 
appropriate to
describe quantum shell effects due to the conduction-band $s$ electrons in noble metals.
For a fuller discussion of the domain of applicability of our continuum approach, 
see Ref.\ \onlinecite{Zhang03}.  
In the present work, Coulomb interactions are included in the self-consistent
Hartree approximation, in order to ensure gauge invariance.

For a system out of equilibrium, there is no general way to define
a thermodynamic free energy. By assuming that the electron motion
is ballistic, however, the energetics of the biased system
can still be described by a nonequilibrium free energy,\cite{Christen97}
which can be used to
study the stability and cohesion of nanowires at finite bias.
We find that metal nanocylinders with certain {\em magic conductance values},
$G/G_0 = 1,3,6,12,17,23,34,42,\ldots$, can support current densities up
to $10^{11} \mbox{A}/\mbox{cm}^2$.  Our finding is consistent with experimental
results for gold nanocontacts 
\cite{Yasuda97,Itakura99,Untiedt00,Hansen00,Yuki01,Mehrez02} ($G < 5 G_0$)
and atomic chains\cite{Agrait02} ($G\simeq G_0$), but implies that the
magic wires with $G>5G_0$ are also extremely robust.
Furthermore, we predict a number of nanowire
geometries that are stable only under an applied bias.

This paper is organized as follows: In Sec.\ \ref{sec:OmegaNeq}, we develop a
formalism to describe the nonequilibrium thermodynamics of a mesoscopic
conductor at finite bias.
In Sec.\ \ref{sec:quasi1D}, we apply this formalism to quasi-one-dimensional
conductors, and obtain gauge-invariant
results for the Hartree potential, grand canonical potential,
and cohesive force of metal nanocylinders at finite bias.
In Sec.\ \ref{sec:Stability},
we perform a linear stability analysis of metal nanocylinders at finite
bias, the principal result of the paper.
Section \ref{sec:conclude} presents some discussion and conclusions.
Details of the stability calculation are presented in Appendix
\ref{append:AA}.

\section{Scattering Approach to Nonequilibrium Thermodynamics }
\label{sec:OmegaNeq}

We consider a metallic mesoscopic conductor 
connected to two reservoirs 
at common temperature $T\equiv (k_B \beta)^{-1}$,
with respective electrochemical potentials $\mu_\pm=\varepsilon_F +eV_\pm$, 
where $\varepsilon_F$ 
is the chemical potential for electrons in the reservoirs at equilibrium, 
$e$ is the electron charge,
and $V_{+(-)}$ is the voltage at the left(right) reservoir.
Because the screening of electric fields in metal nanowires with $G>G_0$ is
quite good, the presence of additional nearby conductors (such as a ground
plane) has a negligible effect on the transport and energetics of 
the system, and is therefore not considered.

While there is no general prescription for constructing a free energy for
such a system out of equilibrium, it is possible to do so based on scattering
theory\cite{Christen97} if inelastic scattering can be neglected, i.e., if the
length $L$ of the conductor satisfies $L \ll L_{\rm in}$.  
In that case, scattering states within the conductor populated by the left (right)
reservoir form a subsystem in equilibrium with that reservoir.
Dissipation only takes place for the outgoing electrons within the reservoir 
where they are absorbed.
Treating electron-electron interactions in mean-field theory, it is then 
possible to
define a nonequilibrium grand canonical potential $\Omega$ of the 
system,
\begin{multline}\label{eq:OmegaNoneq}
  \Omega[\mu_+,\mu_-,U(\rv\,)] =
    \Omega_0[\mu_+,\mu_-,U(\rv\,)] \\
   -\frac12\int\! d^3r \,
    	[\rho_-(\rv\,)+\rho_+(\rv\,)]U(\rv\,),
\end{multline}
where $\Omega_0$ describes independent electrons moving in the
mean field $U$, $\rho_\pm$ are the number densities of electrons ($-$) and
of ionic background charges ($+$), and the second term on the r.h.s.\ of Eq.\
(\ref{eq:OmegaNoneq}) corrects for double-counting of interactions in $\Omega_0$.
Since the electrons injected from the left and right reservoirs
are independent, aside from their
interaction with the mean field, $\Omega_0$ is given by the sum
\begin{equation}\label{eq:Omega0}
  \Omega_0[\mu_+,\mu_-,U(\rv\,)]=\sum_{\alpha=\pm}
  \Omega_\alpha [\mu_\alpha,U(\rv\,)],
\end{equation}
where
\begin{equation}\label{eq:Omegaf}
  \Omega_{\alpha} [\mu_\alpha,U] = - k_B T
       \int\!dE\ g_\alpha (E) \ln\left[1+e^{-\beta(E-\mu_\alpha)}\right]
\end{equation}
is the grand canonical potential of independent electrons
moving in the potential $U$, in equilibrium with reservoir $\alpha$, and
the {\em injectivity}\cite{Buttiker93,Gasparian96,Christen97}
\begin{equation}
g_\alpha(E) = \frac{1}{4\pi i} \sum_\gamma
\mbox{Tr} \left\{S^\dagger_{\gamma\alpha} \frac{\partial S_{\gamma\alpha}}{\partial E}
-\frac{\partial S^\dagger_{\gamma\alpha}}{\partial E} S_{\gamma\alpha}\right\}
\label{eq:pdos}
\end{equation}
is the partial density of states of electrons injected by reservoir $\alpha$.
Here $S_{\gamma\alpha}= S_{\gamma\alpha}[E,U(\rv\,)]$ is the submatrix of the electronic
scattering matrix describing electrons injected from reservoir $\alpha$ and
absorbed by reservoir $\gamma$, and is a functional of the mean-field potential.

The number density $\rho_-(\rv\,)$ of the conduction electrons is
\begin{equation}\label{eq:rhoScreened}
  \rho_-(\rv\,) = \frac{\delta\Omega_0}{\delta U(\rv\,)} 
  = \sum_{\alpha=\pm}\int\!dE\,g_\alpha(\rv,E)f(E,\mu_\alpha),
\end{equation}
where $f(E,\mu)=\left(1+\exp[\beta(E-\mu)]\right)^{-1}$ is the Fermi-Dirac
distribution function, and
\begin{equation}
g_\alpha(\rv,E) = -\frac{1}{4\pi i} \sum_\gamma
\mbox{Tr} \left\{S^\dagger_{\gamma\alpha} \frac{\delta S_{\gamma\alpha}}{\delta U(\rv\,)}
-\frac{\delta S^\dagger_{\gamma\alpha}}{\delta U(\rv\,)} S_{\gamma\alpha}\right\}
\label{eq:lpdos}
\end{equation}
is the {\em local partial density of states}\cite{Buttiker93,Gasparian96,Christen97}
for electrons injected from reservoir $\alpha$.
In Eqs.\ (\ref{eq:rhoScreened}) and (\ref{eq:lpdos}), $\delta/\delta U(\rv\,)$
denotes the functional derivative.

The mean-field potential $U$ is determined in the Hartree approximation by
\begin{equation}
  \label{eq:UH}
  U(\rv\,) = \int\! d^3 r^\prime \,
             V(\rv-\rprime)\delta\rho(\rprime),
\end{equation}
where
$\delta\rho(\rv\,)=\rho_-(\rv\,)-\rho_+(\rv\,)$ is the local 
charge imbalance in the 
conductor and  $V(\rv\,)=e^2/|\rv\,|$ is the Coulomb potential.
The Hartree potential depends on the electrochemical potentials
of the left and right reservoirs.

The whole formalism (\ref{eq:OmegaNoneq})--(\ref{eq:lpdos})
is very similar for any mean-field
potential that is a local functional of the electron density,\cite{Todorov00}
but we choose to work with the Hartree potential for simplicity.
The exchange and correlation contributions to the mean field are taken into
account in the present analysis only macroscopically,\cite{Zhang03}
by fixing the background density $\rho_+$ to its bulk value.
Throughout this paper, we assume $\rho_+=k_F^3/3\pi^2=\mbox{const.}$ 
within the conductor (jellium model).

\section{Quasi-one-dimensional limit}
\label{sec:quasi1D}

Equations (\ref{eq:OmegaNoneq})--(\ref{eq:UH}) provide a set of
equations at finite bias that have to be solved self-consistently.
For a conductor of arbitrary shape,
these equations may be quite difficult to solve.
We therefore restrict our consideration in the following to quasi-one-dimensional
nanoconductors, with axial symmetry about the $z$-axis.
The shape of the conductor is specified by its radius $R(z)$ as a function of $z$,
and we assume $R(z)\ll L$.
For such a quasi-one-dimensional geometry,
we can approximately integrate out the transverse coordinates,
replacing the Coulomb potential by an effective one-dimensional potential
\begin{equation}\label{eq:Coulomb1D}
  V(z,z^{\prime})=\frac{e^2}{[(z-z^{\prime})^2+\frac{\eta}{2}
  \left(R^2(z)+R^2(z^{\prime})\right)]^{1/2}},
\end{equation}
where $\eta$ is a parameter of order unity.
The longitudinal potential $V(z,z^\prime)$ has to be supplemented with
a transverse confinement potential, which we take as a hard wall at the
surface of the wire.\cite{note_hardwall} \nocite{Garcia_Martin96,Yannouleas98,Puska01}
This boundary condition necessitates a careful treatment of surface charges,
as discussed below.
As a consistency check, our final results for the stability and cohesion
are independent of the value of $\eta$
chosen in the effective Coulomb potential.

With this form of the Coulomb potential, the mean field $U(z)$ becomes a function
of the longitudinal coordinate only, 
and Eqs.\ (\ref{eq:OmegaNoneq})--(\ref{eq:UH})
reduce to a series of one-dimensional integral equations, which are much more
tractable.  
The grand canonical potential of a quasi-cylindrical wire of length $L$ 
is
\begin{multline}\label{eq:Omega1D}
  \Omega[\{\mu_\pm\},R(z),U(z)] = \Omega_0[\{\mu_\pm\},R(z),U(z)] \\
    - \frac12\int_0^L\!dz\,[\rho_-(z)+\rho_+(z)]U(z),
\end{multline}
where $\Omega_0$ is still given by
Eqs.\ (\ref{eq:Omega0})--(\ref{eq:pdos}), with
$\Omega_\alpha = \Omega_\alpha[\mu_\alpha,R(z),U(z)]$.  Here
\begin{equation}\label{eq:rhoScreened1D}
  \rho_-(z) = \frac{\delta\Omega_0}{\delta U(z)} 
  = \sum_{\alpha=\pm}\int\!dE\,g_\alpha(z,E)f(E,\mu_\alpha)
\end{equation}
is the linear density of conduction electrons, where
\begin{equation}
g_\alpha(z,E) = -\frac{1}{4\pi i} \sum_\gamma
\mbox{Tr} \left\{S^\dagger_{\gamma\alpha} \frac{\delta S_{\gamma\alpha}}{\delta U(z)}
-\frac{\delta S^\dagger_{\gamma\alpha}}{\delta U(z)} S_{\gamma\alpha}\right\}
\label{eq:lpdos1D}
\end{equation}
is the injectivity of a circular slice of the conductor at $z$.
The scattering matrix $S = S[E,R(z), U(z)]$ is now a functional of $R(z)$ and
$U(z)$.
In order to compensate for the depletion of surface electrons due to the hard-wall
boundary condition, 
the linear density of positive background charges is taken to be
\begin{multline}
\label{eq:rho+}
\rho_{+}(z)=\frac{k_F^3 R(z)^2}{3\pi}
- \frac{k_F^2 R(z)\sqrt{1+R_z^2(z)}}{4} \\
+ \frac{k_F}{3\pi}\left(1-\frac{R(z)R_{zz}(z)}{\sqrt{1+R_z^2(z)}}\right),
\end{multline}
where $R_z(z)=dR(z)/dz$ and $R_{zz}(z)=d^2R(z)/dz^2$.
The second term on the r.h.s.\ of Eq.\ (\ref{eq:rho+})
corresponds to the well-known surface correction in the free-electron
model.\cite{Stafford99}
The last term represents an integrated-curvature contribution,
which is found to be a small correction.
The prescription given in Eq. (\ref{eq:rho+}) is essentially equivalent to the widely employed practice of placing the hard-wall boundary at a distance $d=3\pi/8k_F$ outside the surface of the metal.\cite{Lang73}

The Hartree potential is
\begin{equation}\label{eq:UH1D}
  U(z) = \int_0^L\!dz'\,V(z,z')\delta\rho(z'),
\end{equation}
where $\delta\rho(z) = \rho_-(z) - \rho_+(z)$.

Equations (\ref{eq:Coulomb1D})--(\ref{eq:UH1D}) provide a natural, gauge-invariant,
generalization of the {\em nanoscale free-electron model},\cite{Stafford97a} 
which has been successful in describing many 
equilibrium\cite{Urban04a,Stafford97a} and 
linear-response\cite{Torres94,Burki99,Burki99b}
properties of simple metal nanowires,
to the case of nanowires at finite bias.  
This formalism represents a considerable simplification
compared to {\it ab initio} 
approaches\cite{Todorov00,Ventra01,Brandbyge02,Mehrez02,Mozos02}
or even traditional jellium calculations,\cite{Yannouleas98,Puska01}
and permits analytical results for the cohesion and stability of metal
nanocylinders at finite bias.

\subsection*{Quasicylindrical nanowire without backscattering}

Consider a nearly cylindrical nanowire, with radius 
\begin{equation}\label{eq:Rz}
  R(z) = R_0 + \lambda\deltaR(z), 
\end{equation}
where $\lambda$ is a small parameter and
$\delta R(z)$ is a slowly-varying function.  
The couplings of the nanowire to the reservoirs are assumed ideal, so that electrons
enter or exit the conductor without backscattering.  For sufficiently small $\lambda$,
electron waves partially reflected or transmitted by the small surface modulation
can be neglected, because they give a negligible contribution to 
the density of states.  
The right (left)-moving electrons in the bulk of the wire
are thus in equilibrium with the left (right) reservoir.
Moreover, the Hartree potential $U(z)$ varies slowly with $z$, and 
can be taken as a shift of the conduction-band bottom in the adiabatic approximation.
The injectivity therefore simplifies to 
\[
g_\alpha (z,E) = \frac12 g(z, E-U(z)),
\]
where $g(z,E)$ is the local
density of states for free electrons in a circular slice of radius $R(z)$.
Eq.\ (\ref{eq:Omegaf}) can be rewritten as
\begin{multline}\label{eq:Omegaf1D}
  \Omega_\alpha [\mu,R(z),U(z)] =
    \frac12 \int_0^L\!dz\,\int^{\mu-U(z)}\!dE\,\\
    \times[E-\mu+U(z)]\,g_T(z,E),
\end{multline}
where the integration variable $E$ is no longer the total energy of an electron, but
rather its kinetic energy.
Here
$g_T(z,E)=-\int\!dE'\,g(z,E')\partial f(E-E',0)/\partial E$
is a convoluted local
density of states in a slice of the wire, and can be used to obtain 
finite-temperature thermodynamic quantities from their zero-temperature 
expressions,\cite{Brack97} so that Eq.\ (\ref{eq:Omegaf1D}) is equivalent to the usual 
definition of the grand canonical potential (\ref{eq:Omegaf}).
Similarly, the linear density of electrons can be written in terms of the 
convoluted density of states as
\begin{equation}\label{eq:rho1D}
  \rho_{-}(z) = \frac{\delta\Omega_0}{\delta U(z)}
    = \frac12\sum_{\alpha=\pm}\int^{\mu_\alpha-U(z)}\!\!dE\,g_T(z,E).
\end{equation}

The convoluted density of states of a circular slice of the nanowire can be 
expressed semiclassically as\cite{Brack97}
$g_T(z,E)\equiv \bar{g}_T(z,E)+\delta g_T(z,E)$, where
$\bar{g}_T$ is a smoothly-varying function of the geometry, known as the Weyl term, and $\delta g_T(z,E)$ is an oscillatory quantum correction.
The temperature dependence of the Weyl term is negligible,
$\bar{g}_T=\bar{g}\times\left(1+{\cal O}(T/T_F)^2\right)$, where $T_F$ is the 
Fermi-temperature. The zero-temperature value is
\begin{multline}\label{eq:DOSWeyl}
  \bar{g}(z,E) =
    \frac{k_F^3\partial{\cal V}(z)}{2\pi^2\eps{F}}
      \sqrt{\frac{E}{\eps{F}}} \\
    -\frac{k_F^2\partial{\cal S}(z)}{8\pi\eps{F}}
    +\frac{k_F\partial{\cal C}(z)}{6\pi^2\eps{F}}
      \sqrt{\frac{\eps{F}}{E}},
\end{multline}
where $\partial{\cal V}(z),\,\partial{\cal S}(z)$ and $\partial{\cal C}(z)$ are,
respectively, the volume, external
surface-area, and external mean-curvature of a slice of the wire.
The fluctuating part $\delta g_T$ can be obtained through the trace 
formula\cite{Kassubek01}
\begin{multline}\label{eq:DOSfluct}
  \delta g_T(z,E) = \frac{k_F^2}{2\pi\varepsilon_F}
  \sum_{wv} f_{vw}a_{vw}(T) \\
  \times \frac{L_{vw}(z)}{v^2}\cos\theta_{vw}(z,E),
\end{multline}
where the sum is over all classical periodic orbits $(v,w)$ in a
disk billiard\cite{Balian72,Brack97} of 
radius $R(z)$. Here the factor $f_{vw}=1$ for $v=2w$, and 2 otherwise, accounts for the 
invariance under time-reversal symmetry of some orbits, 
$L_{vw}(z)=2vR(z)\sin(\pi w/v)$ is the length of periodic orbit $(v,w)$,
$\theta_{vw}(z,E)=k_FL_{vw}(z)\sqrt{E/\varepsilon_F}-3v\pi/2$, and 
$a_{vw}(T)=\tau_{vw}/\sinh\tau_{vw}$
is a temperature-dependent damping factor, with $\tau_{vw}=\pi k_FL_{vw}T/2T_F$. 

The semiclassical approximation, Eqs.\ (\ref{eq:DOSWeyl}) and (\ref{eq:DOSfluct}), allows for
an analytical solution for the ballistic nonequilibrium electron distribution in
a metal nanowire at finite bias.  It also enables us to carry out a linear stability analysis
of metal nanowires at finite bias, with analytical results for the stability coefficients.
Although these calculations could in principle be carried out using a fully quantum
mechanical solution of the electronic scattering problem, the
semiclassical approximation has been shown \cite{Urban03,Urban04b} to accurately describe the long-wavelength
surface perturbations that are the limiting factor \cite{Zhang03} in the stability of long nanowires.

Eqs.\ (\ref{eq:UH1D}) and (\ref{eq:rho1D}) provide a set of self-consistent
equations to solve for the ballistic nonequilibrium electron distribution in
a quasi-one-dimensional nanoconductor at finite bias.  Once the distribution $\rho_-$
is obtained, the grand canonical potential $\Omega$ of the electron gas
may be calculated from Eqs.\ (\ref{eq:Omega1D}) and (\ref{eq:Omegaf1D}).  
The functional dependence of 
$\Omega[R(z)]$ yields information on the cohesion\cite{Stafford97a,Stafford99}
and stability\cite{Kassubek01,Zhang03,Urban03,Urban04b} 
of a metal nanowire, as in the equilibrium case.

\subsection*{Solution for a cylindrical nanowire; Hartree potential and tensile force}

\begin{figure}[t]
  \includegraphics[width=0.95\columnwidth]{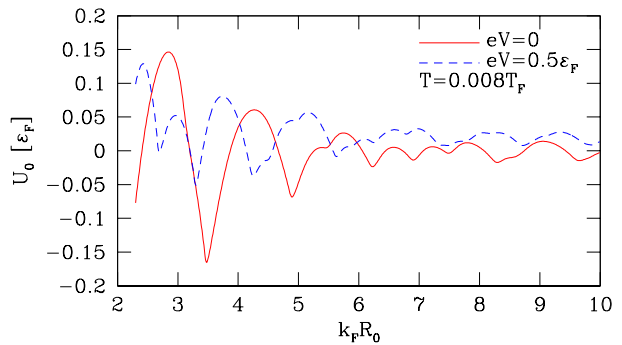}
  \caption{(Color online) The Hartree potential $U_0$ for electrons in
    a cylindrical nanowire at finite temperature $T=0.008\,T_F$,
    with a symmetric potential drop, versus the
    radius  $R_0$ of the wire.
  \label{fig:U0}}
\end{figure}

For an unperturbed cylinder, the mesoscopic Hartree potential $U_0$
that simultaneously solves Eqs.\ (\ref{eq:UH1D}) and (\ref{eq:rho1D})
is only a function of the
radius $R_0$, voltage $eV=\mu_+-\mu_-$, and temperature $T$, and is constant along 
the wire, neglecting boundary effects, which are important only within a
screening length ($\mbox{}\sim k_F^{-1}$) of each contact. 
(This description is valid for wires with $1 \ll k_F L < k_F L_{\rm in}$.)
$U_0$ is independent of the choice of $\eta$ in the
Coulomb interaction, Eq.\ (\ref{eq:Coulomb1D}), and can be determined by the charge neutrality condition
\begin{equation}\label{eq:neutrality}
  Q=\frac12e[N_-(\mu_+-U_0)+N_-(\mu_--U_0)]-eN_+=0,
\end{equation}
where $\frac12 N_-(\mu=\mu_{\pm}-U_0)=\frac12
\int_0^L\!dz\int^\mu\!dE\,g_T(z,E)$ 
is the number of right(left)-moving electrons in the cylindrical wire,
and $N_+$ is the total number of background positive charges.
Equation (\ref{eq:neutrality}) gives a relation\cite{Christen96}
\begin{equation}\label{eq:U0}
  U_0=U_0^{(s)}(R_0,V,T)+\frac12(\mu_++\mu_-)-\varepsilon_F,
\end{equation}
where $U_0^{(s)}$ is calculated with a symmetric voltage drop
$V_+=-V_-=\frac12V$. Equation\ (\ref{eq:U0}) 
guarantees that all physical properties of the system calculated in the
following are just functions of the voltage $V$, and not of
$\mu_+$ and $\mu_-$ separately.

Using these expressions, one can solve Eq.\ (\ref{eq:neutrality}) for
$U_0$ for a symmetric potential drop, $\mu_\pm = \eps{F}\pm\frac12eV$.
The solution is shown in Fig.\ \ref{fig:U0} as a function of radius $R_0$ at 
two different voltages $V$.
In the equilibrium case ($V=0$), $U_0$ oscillates about zero, exhibiting cusps at
the subband thresholds, and increasing in
amplitude as $R_0$ decreases, due to the quantum confinement.  
Note that in equilibrium,
$U_0 \rightarrow 0$ as $R_0 \rightarrow \infty$, consistent with the 
well-known behavior of bulk jellium.  At finite bias, each cusp in $U_0$ splits in two,
corresponding to the {\em subband thresholds for left- and right-moving electrons}:
\begin{equation} \label{eq:threshold}
\mu_{\pm}=\varepsilon_\nu(R_0)+U_0(R_0,V,T), 
\end{equation}
where $\varepsilon_\nu(R_0)$ are the eigenenergies of a disk billiard of radius $R_0$.
This is illustrated in Fig.\ \ref{fig:U0} for $eV=0.5\varepsilon_F$. 
Note that in addition to the splitting, there is a substantial shift of the peak structure at
finite bias.

\begin{figure}[t]
  \includegraphics[width=8cm]{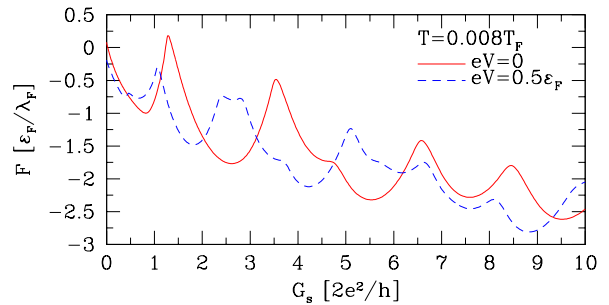}
  \caption{(Color online)
Tensile force in a metal nanocylinder versus cross-sectional area,
    at two different bias voltages at temperature $T=0.008\,T_F$.
    The cross section is plotted in terms of the Sharvin conductance $G_S$,
    Eq.\ (\ref{eq:GS}), and the force is given in units of $\varepsilon_F/\lambda_F$
    ($\lambda_F$ being the Fermi wavelength), which is $1.7$ nN for Au.
  \label{fig:force}
  }
\end{figure}

Using Eq.\ (\ref{eq:Omega1D}), the grand canonical potential of a cylinder is now found to be
\begin{equation}\label{eq:OmegaCyl}
  \Omega(R_0,V,T) = \Omega_0[\{\mu_\pm\},R_0,U_0] - N_+U_0.
\end{equation}
Note that $\Omega$ is invariant under a global shift of the potential $U_0$, due to an
exact cancellation in the two terms on the r.h.s.\ of Eq.\ (\ref{eq:OmegaCyl}).
The tensile force in the nanowire provides direct information about
cohesion, and is given by
\[
F(R_0,V,T)= - \left.\frac{\partial \Omega}{\partial L}\right|_{R_0^2 L, V, T}.
\]
Figure \ref{fig:force} shows the tensile force of a metal nanocylinder
as a function of its cross section for two different bias voltages. 
To facilitate comparison with the stability diagrams 
in Sec.\ \ref{sec:Stability}
below, the cross section
is plotted in terms of the corrected Sharvin conductance\cite{Torres94}
\begin{equation}
G_S = G_0 [(k_F R_0/2)^2 - k_F R_0/2],
\label{eq:GS}
\end{equation}
which gives a semiclassical approximation to the electrical conductance.
In Fig.\ \ref{fig:force}, $F<0$ corresponds to {\em tension}, 
while $F>0$ corresponds to {\em compression}.  As shown below, the cusps
in the cohesive force at the subband thresholds correspond to {\em structural
instabilities} of the system.

In Fig.\ \ref{fig:force}, the force calculated at zero bias is very similar
to previous results\cite{Blom98,Hoeppler99,Stafford99} 
based on the free-electron model, even though those calculations did not
respect the charge neutrality (\ref{eq:neutrality})
enforced by Coulomb interactions.  The reason for the good agreement is that
the contribution of the Hartree potential to the energy of the system
is a {\em second-order} mesoscopic effect at zero bias,\cite{Stafford99,Stafford00} 
which is 
essentially negligible for $G>3G_0$.  An earlier calculation\cite{Ruitenbeek97}
that did invoke charge neutrality obtained a very different---and 
incorrect---result, because the second term on the r.h.s.\ of 
Eq.\ (\ref{eq:OmegaCyl}) was omitted, resulting in a double-counting of
Coulomb interactions.

Figure \ref{fig:force} shows that the cohesive force of a metal nanowire
can be modulated by several nano-Newtons for a bias of a few Volts.
Such a large effect should be observable experimentally using appropriate
cantilevers,\cite{Agrait96,Stalder96,Rubio04} although the intrinsic behavior
might be masked by electrostatic forces in the external circuit.
In contrast to the case at $V=0$, the Coulomb interactions play an essential
role in determining the cohesive force at finite bias, since
the positions of the peaks depend sensitively on the Hartree potential $U_0$ of
the ballistic nonequilibrium electron distribution, shown in
Fig.\ \ref{fig:U0}.  The gauge-invariant result shown in Fig.\ \ref{fig:force}
thus differs substantially from previous results,\cite{Zagoskin98}
where screening was not treated self-consistently.  Gauge-invariant results for the
nonlinear transport through metal nanocylinders will be presented 
elsewhere.\cite{Zhang05}

\section{Linear Stability of a Cylinder at Finite Voltage}
\label{sec:Stability}

In this section, we perform a linear stability analysis\cite{Kassubek01,Zhang03} 
for cylindrical nanowires under a finite bias $V$. Coulomb interactions are
included self-consistently using the formalism of Sec.\ \ref{sec:quasi1D}.
Although the details of the calculation are rather complicated, the method is conceptually
straightforward:
Having found the self-consistent solution (\ref{eq:OmegaCyl})
for a cylinder, we perturb the cylinder as in 
Eq.\ (\ref{eq:Rz}), and expand the free energy up to second order in the small 
parameter $\lambda$.
First, the self-consistent integral equations (\ref{eq:UH1D}) and (\ref{eq:rho1D})
for the ballistic nonequilibrium electron distribution
are solved using first-order perturbation theory in $\lambda$.
Then the free energy is calculated using Eqs.\ (\ref{eq:Omega1D}) and (\ref{eq:Omegaf1D}).

The radius of the wire is given by Eq.\ (\ref{eq:Rz}), with a perturbation function
\[
  \deltaR(z)=\sum_{q}b(q)e^{iqz},\quad
  \quad b(q)^*=b(-q).
\]
The surface perturbation $\lambda \delta R(z)$ is subject to a constraint fixing
the total number of atoms in the wire.  In previous works, we have considered
various constraints,\cite{Stafford99,Urban04b} which allow one to adjust the surface
properties of the wire to model various materials.
The simplest such constraint is volume conservation, under which the coefficient $b(0)$ is fixed by
\[
  b(0) = -\frac{\lambda}{2R_0}\sum_q|b(q)|^2.
\]
Other reasonable constraints do not lead to qualitatively different conclusions.

Within linear response theory, we can expand $\delta\rho(z)$
around $U_0$ to linear order in $U(z)-U_0$, where
$U_0$ is the mesoscopic Hartree potential for the corresponding
unperturbed cylindrical wire. One gets
\begin{multline}\label{eq:rhoLinear}
  \delta\rho(z) \simeq \delta\rho_0(z)-\int\!dz^{\prime}
    \left.\frac{\delta\rho_-(z)}{\delta U(z^{\prime})}\right|_{U_0}U_0
     \\
  + \int\!dz_1dz_2 \left.\frac{\delta\rho_-(z)}{\delta U(z_1)}
      \right|_{U_0}V(z_1,z_2)\delta\rho(z_2),
\end{multline}
where Eq.~(\ref{eq:UH1D}) has been used and $\delta\rho_0(z)$,
called the bare charge imbalance, is defined as
\begin{equation}\label{eq:deltarho0}
  \delta\rho_0(z) =
    \frac12\sum_{\alpha=\pm}\int^{\mu_\alpha-U_0}\!\!dE \, g_T(z,E) - \rho_+(z).
\end{equation}
Now defining the dielectric function
\begin{equation}\label{eq:dielectric1D}
  \epsilon(z_1,z_2) = \delta(z_1-z_2)
    - \int dz_3\left.\frac{\delta\rho_-(z_1)}{\delta U(z_3)}
    \right|_{U_0}V(z_3,z_2),
\end{equation}
we can rewrite Eq.\ (\ref{eq:rhoLinear}) as
\begin{equation}\label{eq:deltarho1D}
  \delta\rho(z) = \int\!dz^{\prime}\epsilon^{-1}(z,z^{\prime})
  \delta\bar{\rho}_0(z^{\prime}),
\end{equation}
where $\epsilon^{-1}$ is the inverse dielectric function which
satisfies $\int dz_3\epsilon^{-1}(z_1,z_3)\epsilon(z_3,z_2)=\delta(z_1-z_2)$,
and
\begin{equation}\label{eq:rhobar}
  \delta\bar{\rho}_0(z) = \delta\rho_0(z)
  -\int\!dz_1\left.\frac{\delta\rho_-(z)}{\delta U(z_1)}\right|_{U_0}U_0.
\end{equation}
The functional derivative 
$\delta\rho_-(z)/\delta U(z^{\prime})$
can be calculated using Eq.\ (\ref{eq:rho1D}), and is found to be
\[
  \left.\frac{\delta\rho_-(z)}{\delta U(z^{\prime})}\right|_{U_0}
    = -\frac{\delta(z-z')}2\sum_{\alpha=\pm}g_T(z,\mu_\alpha-U_0).
\]

Now we can expand the nonequilibrium grand canonical potential~(\ref{eq:Omega1D}) as a
series in $\lambda$.
In order to do so, we first expand
Eq.~(\ref{eq:Omega1D}) around $U_0$. Using Eqs.~(\ref{eq:UH1D}),
(\ref{eq:rho1D}), (\ref{eq:deltarho0}) and (\ref{eq:rhobar}),
one gets
\begin{multline}\label{eq:Omega1DexpandA}
  \Omega=\Omega_0[\{\mu_\pm\},R(z),U_0]
    - U_0\int_0^L\!dz\,\delta\bar{\rho}_0(z) \\
    + \frac12\int_0^L\!dz\,dz^{\prime}\,\delta\bar{\rho}_0(z)\Vsc(z,z^{\prime})
      \delta\bar{\rho}_0(z^{\prime}) \\
    + \frac{(U_0)^2}{4}\sum_{\alpha=\pm}
      \int_0^L\!dz\,g_T(z,\varepsilon_\alpha)
    - N_+U_0,
\end{multline}
where $\varepsilon_{\pm}=\mu_\pm-U_0$, and the
screened potential $\Vsc (z,z^{\prime})$ is defined as
\begin{equation}\label{eq:screenedV}
  \Vsc(z,z^{\prime})=\int dz_1V(z,z_1)\epsilon^{-1}(z_1,z^{\prime}).
\end{equation}
\begin{figure}[b]
\begin{center}
  \includegraphics[width=0.95\columnwidth,draft=false]{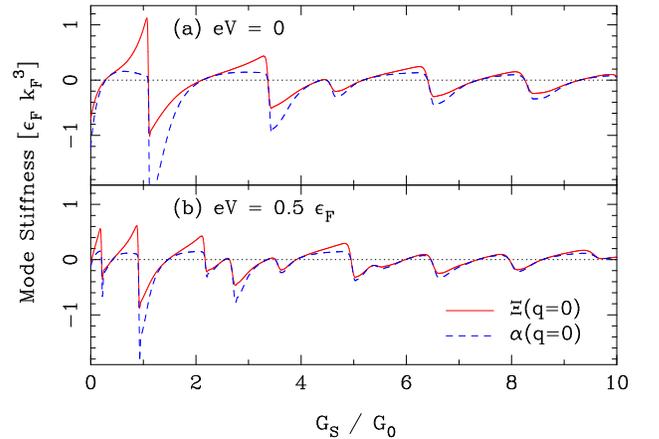}
  \caption{(Color online)
  	Total mode stiffness $\Xi(q=0;R_0,V,T)$, and leading-order contribution
	$\alpha(q=0;R_0,V,T)$, at zero and finite bias, for $T=0.008\,T_F$.
  }
  \label{fig:stiffness_v}
\end{center}
\end{figure}

At this point, all quantities have been expressed in terms of
the local density of states $g_T(z,E)$ and 
the Coulomb interaction $V(z,z^{\prime})$, whose expansions in series of $\lambda$ 
are presented in Appendix \ref{append:AA}.
In the end, the expansion of the grand canonical potential
as a series in $\lambda$ is found to be
\begin{equation}\label{eq:Omega1DexpandB}
  \Omega = \Omega(R_0,V,T) + \lambda^2 L
    \sum_{q>0}\Xi(q;R_0,V,T)|b(q)|^2
\end{equation}
plus terms ${\cal O}(\lambda^3,L^0)$, where
$\Omega(R_0,V,T)$ is given by Eq.\ (\ref{eq:OmegaCyl}), and the mode stiffness
\begin{multline}\label{eq:stabcoef}
  \Xi(q;R_0,V,T) \equiv \alpha(q;R_0,V,T) \\
  + \text{Re}\bigg(\frac{\Vq{q}}{\epsq{q}}\bigg)
    (\varrho^{(1)}_1-\varrho^{(1)}_2q^2)^2.
\end{multline}
Here $\Vq{q}$ and $\epsq{q}$ are, respectively, the Fourier transforms of
$\Vo{z}$ and $\epso{z}$,
the Coulomb potential and dielectric function of an unperturbed cylinder.
The factors $\varrho^{(1)}_1$ and
$\varrho^{(1)}_2$ are given in Eqs.\ (\ref{eq:A1}) and\ (\ref{eq:A2}).
The factor $\alpha(q;R_0,V,T)$ comes from the expansion of $\Omega_0[\{\mu_\pm\},R(z),U_0]$
and is found to be
\begin{multline} \label{eq:alpha}
  \alpha(q;R_0,V,T) = - \frac{\pi\sigma_s}{R_0}
    \frac{\eps+^2+\eps-^2}{\eps{F}^2} \\
  + \left(\pi\sigma_sR_0\frac{\eps+^2+\eps-^2}{\eps{F}^2}
      -\gamma_s \frac{\eps+^{3/2}+\eps-^{3/2}}{\eps{F}^{3/2}}
    \right) q^2 \\
  + \left(
  	\frac{\partial^2}{\partial R_0^2}
	- \frac{1}{R_0}\frac{\partial}{\partial R_0}\right) V_{\rm shell}(R_0,V,T),
\end{multline}
where $\sigma_s$ is the surface tension, 
$\gamma_s$ is the curvature energy, 
and $V_{\rm shell}(R_0,V,T)$ is the mesoscopic {\em electron-shell potential},\cite{Burki03}
given self-consistently at finite bias by
\begin{multline}\label{eq:Vshell}
  V_{\rm shell}(R_0,V,T) = \frac{1}{\pi\eps{F}}
    \sum_{\alpha=\pm}\sum_{wv} \frac{a_{vw}(T)f_{vw}}{v^2L_{vw}} \\
    \times\eps\alpha^2\cos(k_{F}^\alpha L_{vw}-3v\pi/2),
\end{multline}
where $k_F^{\pm}=k_F\sqrt{\eps{\pm}/\eps{F}}$.
In the present article, the values\cite{Stafford97a}
$\sigma_s=\eps{F}k_F^2/16\pi$ and $\gamma_s=2\eps{F}k_F/9\pi^2$, appropriate for a
constant-volume constraint, are used throughout.
Inserting material-specific values\cite{Zhang03} does not lead to a significant
change in the stability diagram.
Equations (\ref{eq:Omega1DexpandB})--(\ref{eq:Vshell}) represent the central result of this paper.

Figure \ref{fig:stiffness_v} shows the long-wavelength mode stiffness $\Xi(q=0)$ at zero and finite bias.
For comparison, the leading-order contribution $\alpha(q=0)$ is plotted as a dashed curve.
The second term on the r.h.s.\ of Eq.\ (\ref{eq:stabcoef}), which is second-order in the 
induced charge imbalance, gives a significant contribution for small radii, but is negligible for
$k_F R_0 \gg 1$.  Moreover, the sign of $\Xi$, which determines stability, is essentially fixed
by $\alpha$ alone.  The relative unimportance of the second-order correction is reminiscent of 
the {\em Strutinsky theorem}\cite{Strut68,Ullmo01} for finite fermion systems, 
which states that shell effects are dominated by the
single-particle contribution in the mean-field potential.

\subsection*{Stability Diagram}

The stability of a cylindrical nanowire of radius $R_0$ at bias $V$ and temperature $T$
is determined by the function $\Xi(q;R_0,V,T)$:  If $\Xi(q)>0$ $\forall q$, then the
nanowire is stable with respect to small perturbations, and is a (meta)stable thermodynamic
state.
If $\Xi(q)<0$ for any $q$, then the wire is unstable.

\begin{figure}[t]
\begin{center}
  \includegraphics[width=0.95\columnwidth,draft=false]{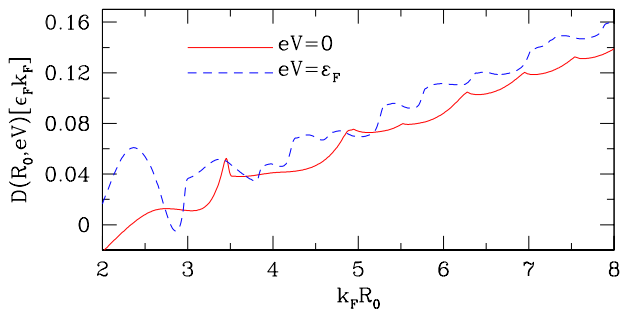}
  \caption{(Color online)
The factor $D(R_0,V)$, coefficient of the $q^2$ contribution
    to $\Xi(q;R_0,V)$, at temperature $T=0.008\,T_F$.}
  \label{fig:alpha_D}
\end{center}
\end{figure}

\begin{figure}[t]
\begin{center}
  \includegraphics[width=0.95\columnwidth,draft=false]{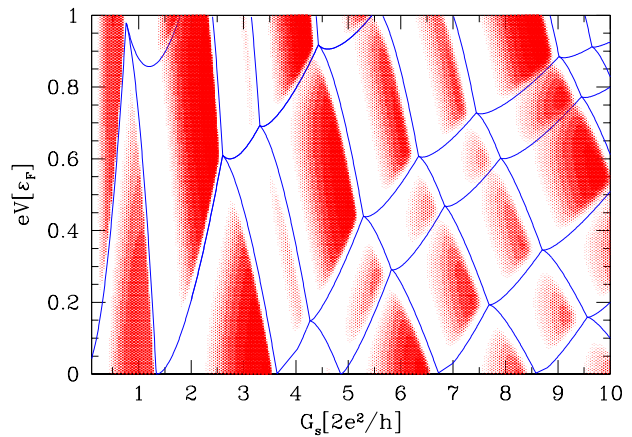}
  \caption{(Color online)
 Stability of cylindrical metal nanowires versus Sharvin
          conductance (\ref{eq:GS}) and bias voltage.
	  Shaded (red) areas indicate stability with respect to small
	  perturbations at $T=0.008\,T_F$.
	  Solid lines indicate subband thresholds for right- and
	  left-moving electrons.}
  \label{fig:stability_v}
\end{center}
\end{figure}

\begin{figure}[b]
\begin{center}
  \includegraphics[width=0.95\columnwidth,draft=false]{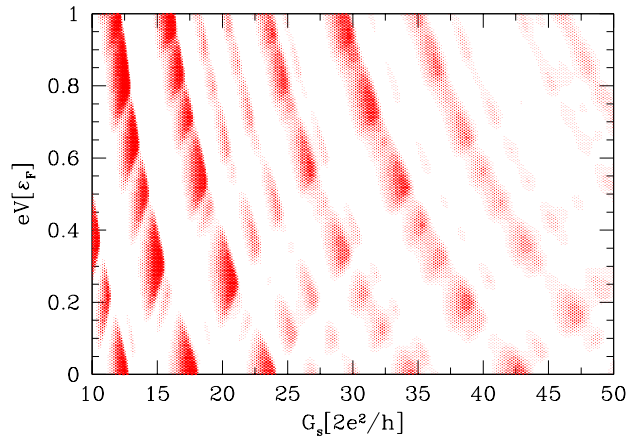}
  \caption{(Color online)
 Stability of cylindrical metal nanowires versus Sharvin conductance
          $G_S$ between $10$ and $50 G_0$ and bias voltage.
	  Shaded (red) areas indicate stability with respect to small
	  perturbations at $T=0.008\,T_F$.
  }
  \label{fig:stability_v2}
\end{center}
\end{figure}

The second term on the r.h.s.\ of Eq.\ (\ref{eq:stabcoef}) is positive semidefinite, and thus cannot
lead to an instability.  
The first term $\alpha(q)$ describes instabilities in two different regimes,
as in the equilibrium case:\cite{Kassubek01,Zhang03} 
({\it i}) The electron-shell contribution has deep negative peaks
at the thresholds to open new conducting subbands (c.f.\ Fig.\ \ref{fig:stiffness_v}).
({\it ii}) The surface contribution to $\alpha(q)$ becomes negative for $qR_0 <1$, the classical Rayleigh instability.
From Eqs.\ (\ref{eq:stabcoef}) and (\ref{eq:alpha}), it is apparent that the most unstable mode (if any) within the semiclassical approximation is $q=0$, except for unphysical radii
$R_0 \lesssim \gamma_s/\sigma_s$ (less than one atom thick).   
To illustrate this point, the second derivative 
$D(R_0,V,T)\equiv\left.\partial^2\Xi/\partial q^2\right|_{q=0}$ 
is shown in Fig.~\ref{fig:alpha_D}.
Note that $D>0$ for $k_F R_0 > 3$.

The stability properties of the system are thus completely determined\cite{note_Peierls}
by the sign of
the stability function $A(R_0,V,T)\equiv\Xi(q=0;R_0,V,T)$.
Figure\ \ref{fig:stability_v} shows a stability diagram in the voltage $V$ and radius $R_0$ plane.
The $x$-axis is given in terms of the Sharvin conductance (\ref{eq:GS}), to facilitate the identification
of the quantized (linear-response) conductance values of the stable nanowires.
The shaded regions show nanowires that are stable with respect to small perturbations, with darker
regions representing larger values of $A(R_0,V,T)$.
In the figure, the solid lines show the subband thresholds for right- and left-moving electrons, which
are determined by Eq.\ (\ref{eq:threshold}).
At the temperature shown $T=0.008T_F$, which corresponds roughly to room temperature, the electron-shell
effect dominates, leading to instabilities at the subband thresholds, and stabilizing the wire in
some of the intervening fingers.

A stability diagram up to $G_S=50 G_0$ is shown in Fig.\
\ref{fig:stability_v2}, where the subband thresholds have been omitted to avoid clutter.
Figures \ref{fig:stability_v} and \ref{fig:stability_v2} show that cylindrical metal nanowires with
certain {\em magic conductance values} $G/G_0=1,3,6,12,17,23,34,42,\ldots$ remain linearly stable at
room temperature up to bias voltages $eV \sim 0.1 \varepsilon_F$ or higher.
These ballistic conductors can therefore support
extremely high current densities, of order $10^{11}\mbox{A}/\mbox{cm}^2$ by Eq.\ (\ref{eq:bigj}).
These are precisely the same magic cylinders which were previously
found to be linearly stable at zero bias up to very high temperatures.\cite{Zhang03}
Cylinders with $G/G_0=8$ and 10 are also predicted to be stable at finite bias, but not as
robust as the neighboring configurations with $G/G_0=6$ and 12.  

It should be mentioned that in addition to the stable cylindrical configurations shown in
Figs.\ \ref{fig:stability_v} and \ref{fig:stability_v2}, nanowires with elliptical
cross sections and conductance
$G/G_0=2,5,9,29,\ldots$ were also found to be stable at zero bias,\cite{Urban04b}
although their finite-bias stability has not yet been investigated.

Perhaps the most startling prediction of Figs.\ \ref{fig:stability_v} and \ref{fig:stability_v2} is that
there are a number of cylindrical nanowire structures which are stable with respect to small perturbations
at finite bias, but unstable in equilibrium!   These metastable structures could lead to additional
peaks in conductance histograms at finite bias, which are not present at low bias.
It may also be possible to observe switching behavior between the various stable structures as
the voltage is varied.

The results of the above
stability analysis should be directly relevant for nanowires made of simple
monovalent metals, such as alkali metals and, to some extent, noble metals.  Indeed,
the calculated bias dependence of the stability of metal nanocylinders with conductance
$G/G_0=1$ and 3, shown in Fig.\ \ref{fig:stability_v}, is consistent with experimental histograms
for gold nanocontacts,\cite{Yasuda97} where a peak at $G\approx G_0$ was found up to
1.9V at room temperature, and a peak at $G\approx 3 G_0$ was found up to about 1.5V.
Similar experimental results have been obtained by several
groups.\cite{Itakura99,Untiedt00,Hansen00,Yuki01,Mehrez02,Agrait02}
Our analysis strongly suggests that the remarkable stability properties of gold nanowires at
finite bias are not a special property of gold, but rather a generic feature of metal nanoconductors.

\subsection*{Nature of the instability}
\label{sec:electromigration}

The nature of the predicted instability of metal nanocylinders at finite bias may be illuminated
by means of a nontrivial identity\cite{Zhang03} linking Eqs.\ (\ref{eq:OmegaCyl}) and
(\ref{eq:stabcoef}):
\begin{equation} \label{eq:q0id}
 \lim_{q\rightarrow 0} \Xi(q;R_0,V,T) =
    \left(\frac{\partial^2}{\partial R_0^2}
-\frac{1}{R_0}\frac{\partial}{\partial R_0}\right)
    \frac{\Omega(R_0,V,T)}{L}.
\end{equation}
This implies that the instability corresponds to a {\em homogeneous-inhomogeneous transition},\cite{Zhang03}
since the r.h.s.\ of Eq.\ (\ref{eq:q0id}) is proportional to the energetic cost of a volume-conserving
phase separation into thick and thin segments.  In the inhomogeneous phase at finite bias, the
surface corrugation will not be static, but will diffuse like a defect undergoing 
{\em electromigration}.\cite{Yasuda97,Ralls89,Holweg92}
The stable nanocylinders are immune to electromigration, because they are translationally invariant and
they are so thin that they are defect free.  Electromigration is possible only if a surface-defect
is nucleated,\cite{Burki04} which becomes energetically favorable on the stability boundary.
The predicted surface instability may thus represent the ultimate nanoscale limit of electromigration.

\section{Conclusions}
\label{sec:conclude}

In this paper, we develop a self-consistent scattering approach to the
nonequilibrium thermodynamics of open mesoscopic systems, and use it to study
the cohesion and stability
of metal nanocylinders under finite bias.  In our approach,
the positive ions are modeled as an incompressible
fluid, and interactions are treated in the Hartree approximation, using a 
quasi-one-dimensional form of the Coulomb interaction.  This single-band model is appropriate for
simple monovalent metals.  It is especially suited to alkali metals, but is also appropriate to
describe quantum shell effects due to the conduction-band $s$ electrons in noble metals.

We have utilized a {\em semiclassical} treatment of the electron-shell structure that plays a 
crucial role in stabilizing metal nanowires at finite bias.  Previous studies
\cite{Urban03,Urban04b} have shown that this semiclassical approach accurately describes the
energetic cost of long-wavelength surface perturbations, which are the limiting factor \cite{Zhang03}
in the structural stability of long nanowires.  Furthermore, we have assumed a {\em ballistic
nonequilibrium electron distribution} in the nanowire at finite bias, 
neglecting inelastic electron-phonon and electron-electron scattering.  This approximation is valid
for wires shorter than the inelastic mean-free path.

We find that the tensile force in a nanowire can be modulated by several nano-Newtons 
when biased by a few volts. Such a large effect should be observable 
experimentally,\cite{Agrait96,Stalder96,Rubio04} although the intrinsic behavior might be 
masked by electrostatic forces in the external circuit.
  
The principal result of this paper is a linear stability analysis of metal nanowires at
finite bias, which reveals that cylindrical wires with certain magic conductance
values $G/G_0 = 1,3,6,12,17,23,34,42,\ldots$ remain stable up to bias voltages 
$eV\sim 0.1\,\varepsilon_F$ or higher,
with the maximum sustainable bias decreasing with increasing radius.
In particular, wires with $G/G_0 =1$ and $3$ are predicted to be stable up to
$eV\sim 0.5\,\varepsilon_F$.
This maximum voltage is slightly larger than what is observed experimentally.\cite{Yasuda97,Agrait02}
It should, however, be pointed out that stability with respect
to small perturbations is not a sufficient condition for a nanowire to be observed:
Metal nanowires are metastable structures, and can be observed only if their {\em lifetime}
is sufficiently long on the experimental timescale.
As a result, the observed maximum sustainable bias is 
likely to be somewhat smaller than that predicted by a linear stability analysis.

A striking prediction of our stability analysis is the existence of nanowire structures 
(e.g.\ cylinders with conductance $G/G_0=2,5,7,9,14,20,\ldots$)
that are only stable under an applied bias. 
This suggests that conductance histograms taken at finite voltage might have additional peaks, 
or even a completely different set of peaks, compared to zero-voltage histograms.  
It may also be possible to observe switching between different stable structures as a 
function of voltage.

Metal nanowires with elliptical cross sections and conductance $G/G_0=2,5,9,29,\ldots$ are also
predicted to be stable at zero bias.\cite{Urban04b}  Although some of the conductance values
of the elliptical wires coincide with those of cylindrical wires predicted to be stable only 
at finite bias, it should be possible to distinguish these geometries experimentally due to the
different kinetic pathways involved in their formation, and the very different bias dependence
of their stability.

Finally, we point out that the predicted instability of metal nanowires at finite bias may 
represent the 
ultimate nanoscale limit of {\em electromigration}, due to the current-induced nucleation of
surface modulation in an otherwise perfect, translationally invariant nanowire.

\begin{acknowledgments}
This work was supported by NSF Grant No.\ 0312028.  CAS thanks Hermann Grabert
and Frank Kassubek for useful discussions during the early stages of this work.
\end{acknowledgments}

\appendix
\section{Expansion of the Non-Equilibrium Free Energy}
\label{append:AA}

In this appendix, we present more details of the derivation of Eqs.~(\ref{eq:Omega1DexpandA})
and (\ref{eq:Omega1DexpandB}).

\subsection*{Local Density of States $g_T(z,E)$}\label{app:dos}

In order to include the temperature
in the semiclassical formalism, we use a convoluted density of
states $g_T(z,E)=\int\left(-\partial f_0(E-E')/\partial E\right)g(z,E')dE'$, where $f_0(E)=[1+\exp(\beta E)]^{-1}$. Thermodynamic quantities are then obtained through their zero-temperature expression with the density of states $g(E)$ replaced by $g_T(E)$.
The temperature dependence of the average part $\bar{g}_T(z,E)$, proportional
to $(1+{\cal O}(T/T_F)^2) \approx 1$, ($T_F$ is the Fermi temperature), is negligible, while the temperature dependence of the fluctuating part $\delta g_T(z,E)$ is included in the damping factor $a_{vw}(T)$ [see Eq.\ (\ref{eq:DOSfluct})].
In the following, we set the factor $a_{vw}(T)$ equal to its unperturbed
value at the Fermi energy $\varepsilon_F$, since the variation of this factor with the perturbation or with energy does not give an important contribution.
We also drop the subscript $T$ for the convoluted density of states to simplify the notation.

For a perturbed cylinder, the average part of the density of states, Eq.\ (\ref{eq:DOSWeyl}), can be expanded to second order in the small parameter $\lambda$ as
\begin{equation}\label{eq:dosWeylexpand}
  \bar{g}(z,E) \simeq \bar{g}^{(0)} + \lambda\bar{g}^{(1)}
  + \lambda^2\bar{g}^{(2)},
\end{equation}
where
\begin{eqnarray*}
  \bar{g}^{(0)} &=&
    \frac{k_F^3R_0^2}{2\pi \varepsilon_F}\sqrt{\frac{E}{\varepsilon_F}}
    - \frac{k_F^2R_0}{4\varepsilon_F}
    + \frac{k_F}{6\pi\varepsilon_F}\sqrt{\frac{\varepsilon_F}{E}}, \\
  \bar{g}^{(1)} &=&
    \left(\frac{k_F^3R_0}{\pi\varepsilon_F}
      \sqrt{\frac{E}{\varepsilon_F}}
    - \frac{k_F^2}{4\varepsilon_F}\right)\deltaR(z) \\
    &&- \frac{k_FR_0}{6\pi\varepsilon_F}
      \sqrt{\frac{\varepsilon_F}{E}}\deltaR^{\prime\prime}(z), \\
  \bar{g}^{(2)} &=&\frac{k_F^3}{2\pi\varepsilon_F}\sqrt{\frac{E}
    {\varepsilon_F}}\deltaR^2(z)
    -\frac{k_F^2R_0}{8\varepsilon_F}
      \deltaR^{\prime\,2}(z) \\
    &&- \frac{k_F}{6\pi\varepsilon_F}\sqrt{\frac{\varepsilon_F}{E}}
      \deltaR(z)\deltaR^{\prime\prime}(z),
\end{eqnarray*}
and the prime denotes differentiation with respect to $z$.

Similarly, the fluctuating part of the density of states for a small deformation of a cylinder, Eq.\ (\ref{eq:DOSfluct}), can be calculated using semiclassical perturbation theory,\cite{Kassubek01,Stafford01} and is found to be
\begin{equation}\label{eq:ddosexpand}
  \delta g(z,E) \simeq \delta g^{(0)}(E) + \lambda\delta g^{(1)}(z,E)
  + \lambda^2\delta g^{(2)}(z,E),
\end{equation}
with
\begin{equation*}
  \delta g^{(0)} = \frac{k_F^2}{2\pi\varepsilon_F}
  \sum_{wv} \frac{a_{vw}(T)f_{vw}L_{vw}}{v^2}\cos\theta_{vw}(E),
\end{equation*}
\begin{multline*}
  \delta g^{(1)} = \frac{k_F^2}{2\pi\varepsilon_F}
    \sum_{wv}\frac{a_{vw}(T)f_{vw}L_{vw}}{v^2R_0}\deltaR(z) \\
  \times \left[\cos\theta_{vw}(E) - k_EL_{vw}\sin\theta_{vw}(E)\right],
\end{multline*}
\begin{multline*}
  \delta g^{(2)} = -\frac{k_F^2}{2\pi\varepsilon_F}
    \sum_{wv}\frac{a_{vw}(T)f_{vw}L_{vw}^2}{2v^2R_0^2}k_E\deltaR^2(z) \\
  \times \left[2\sin\theta_{vw}(E) + k_EL_{vw}\cos\theta_{vw}(E)\right].
\end{multline*}
 where, once more, the sum is over all classical periodic orbits $(v,w)$ in a disk billiard of radius $R_0$, the factor $f_{vw}=1$ for $v=2w$ and 2 otherwise accounts for the invariance under time-reversal symmetry of some orbits, $L_{vw}=2vR_0\sin(\pi w/v)$ is the length of periodic orbit $(v,w)$,
$\theta_{vw}(E)=k_FL_{vw}\sqrt{E/\varepsilon_F}-3v\pi/2$, and $a_{vw}(T)=\tau_{vw}/\sinh\tau_{vw}$, with $\tau_{vw}=\pi k_FL_{vw}T/2T_F$, is a temperature dependent damping factor.

To shorten subsequent equations, we define the functions of energy $g^{(i)}_j(E)$ by writing the first- and second-order contributions to the local density of states as
\begin{align}\label{eq:ldos_expansion}
  g^{(1)}(z,E) &= g^{(1)}_1(E)\deltaR(z)
  	+ g^{(1)}_2(E)\deltaR^{\prime\prime}(z), \\
  g^{(2)}(z,E) &= g^{(2)}_1(E)\deltaR^2(z)+g^{(2)}_2(E)\deltaR^{\prime2}(z) \notag \\
  	       &\quad + g^{(2)}_3(E)\deltaR(z)\deltaR^{\prime\prime}(z).
\end{align}
The total density of states per unit length of a cylindrical wire is $g^{(0)}(E)=\bar{g}^{(0)}(E)+\delta g^{(0)}(E)$.

\subsection*{Bare charge imbalances $\delta\rho_0(z)$
		and $\delta\bar{\rho}_0(z)$}\label{app:drho}

Now substituting the expansion of the local density of states 
into Eq.~(\ref{eq:deltarho0}), one gets the
expansion of $\delta\rho_0(z)$ as 
\begin{equation}\label{eq:drhoexpand}
  \delta \rho_0(z) \simeq \lambda\delta\rho_0^{(1)}
    + \lambda^2\delta\rho_0^{(2)},
\end{equation}
with
\begin{equation*}
  \delta\rho^{(1)}_0(z) = \varrho^{(1)}_1\deltaR(z) + \varrho^{(1)}_2\deltaR^{\prime\prime}(z),
\end{equation*}
where the coefficients $\varrho^{(1)}_1$ and $\varrho^{(1)}_2$ are defined as
\begin{multline}\label{eq:A1}
  \varrho^{(1)}_1 = \frac{k_F^3R_0}{3\pi}\left(
    \frac{\eps+^{3/2}+\eps-^{3/2}}{\eps{F}^{3/2}} - 2\right) \\
  -\frac{k_F^2}{8}\left(\frac{\eps++\eps-}{\eps{F}}-2\right)
    + \frac{\delta \varrho^{(1)}_++\delta \varrho^{(1)}_-}{2},
\end{multline}
\begin{multline}\label{eq:A2}
  \varrho^{(1)}_2 = -\frac{k_FR_0}{6\pi}
    \left(
    \frac{\eps+^{1/2}+\eps-^{1/2}}{\eps{F}^{1/2}}-2\right)
    , \hfill
\end{multline}
with $\eps\pm=\mu_\pm - U_0$, and
\begin{multline*}
  \delta \varrho^{(1)}_{\pm} = \frac{k_F^2\eps{\pm}}{\pi\eps{F}}
    \sum_{vw}\frac{f_{vw}a_{vw}(T)}{v^2R_0} \\
    \times\bigg(L_{vw}\cos\theta_{vw}(\varepsilon_{\pm})
  + k_\pm^{-1}\sin\theta_{vw}(\varepsilon_{\pm})\bigg),
\end{multline*}
where $k_\pm=k_F\sqrt{\eps\pm/\eps{F}}$, while
\begin{equation*}
  \delta\rho^{(2)}_0(z) = \varrho^{(2)}_1\deltaR^2(z) 
+ \varrho^{(2)}_2\deltaR^{\prime\,2}(z)
    + \varrho^{(2)}_3\deltaR(z)\deltaR^{\prime\prime}(z),
\end{equation*}
where coefficients $\varrho^{(2)}_1, \varrho^{(2)}_2$ and $\varrho^{(2)}_3$ are
\begin{align*}
  \varrho^{(2)}_1 &=
     \frac{k_F^3}{6\pi}\left(
     \frac{\eps+^{3/2}+\eps-^{3/2}}{\eps{F}^{3/2}} - 2\right)
    + \frac{\delta \varrho^{(2)}_++\delta \varrho^{(2)}_-}{2}, \\
  \varrho^{(2)}_2 &= -\frac{k_F^2R_0}{16}\left(\frac{\eps++\eps-}{\eps{F}}-2\right), \quad
  \varrho^{(2)}_3 = \frac{\varrho^{(1)}_2}{R_0},
\end{align*}
and
\begin{multline*}
  \delta \varrho^{(2)}_{\pm} = -\frac{k_F^2}{2\pi}
    \frac{\eps{\pm}}{\eps{F}}
    \sum_{vw}\frac{f_{vw}a_{vw}(T)}{v^2R_0^2}L_{vw} \\
  \times\bigg(L_{vw}\sin\theta_{vw}(\eps{\pm})
    - 2k_\pm^{-1}\cos\theta_{vw}(\eps{\pm})\bigg).
\end{multline*}
Similarly, Eq.\ (\ref{eq:rhobar}) for $\delta\bar{\rho}_0(z)$ is expanded as
\begin{equation}\label{eq:drhobarexpand}
  \delta \bar{\rho}_0(z,E) \simeq \delta\bar{\rho}_0^{(0)} +
    \lambda\delta\bar{\rho}_0^{(1)}
    + \lambda^2\delta\bar{\rho}_0^{(2)},
\end{equation}
with
\begin{eqnarray*}
  \delta\bar{\rho}^{(0)}_0 &=
    &\frac{U_0}{2}\left[g^{(0)}(\eps+)+g^{(0)}(\eps-)\right], \\
  \delta\bar{\rho}^{(1)}_0 &=
    & \bar\varrho^{(1)}_1\deltaR(z)
    + \bar\varrho^{(1)}_2\deltaR^{\prime\prime}(z), \\
  \delta\bar{\rho}^{(2)}_0 &=
    & \bar\varrho^{(2)}_1\deltaR^2(z)
      + \bar\varrho^{(2)}_2\deltaR^{\prime2}(z)
      + \bar\varrho^{(2)}_3\deltaR(z)\deltaR^{\prime\prime}(z),
\end{eqnarray*}
where $\bar\varrho^{(i)}_j \equiv \varrho^{(i)}_j
  + \frac12U_0[g^{(i)}_j(\varepsilon_+)+g^{(i)}_j(\varepsilon_-)]$.

\subsection*{Effective Coulomb potential $V(z,z^{\prime})$}\label{app:V}

The expansion of the Coulomb potential (\ref{eq:Coulomb1D}) as a series in $\lambda$ gives
\begin{multline}\label{eq:Vexpand}
  V(z,z') \simeq \Vo{z-z'} \\
    + \lambda\,V^{(1)}(z,z') + \lambda^2\,V^{(2)}(z,z'),
\end{multline}
where
\begin{align*}
  \Vo{z} &=
    \frac{e^2}{\sqrt{z^2+\eta R_0^2}}, \\
  V^{(1)}(z,z^{\prime}) &= \frac{\deltaR(z)+\deltaR(z^{\prime})}{2}
    \cdot\frac{d\Vo{z-z^{\prime}}}{dR_0}, \\
  V^{(2)}(z,z^{\prime}) &= \frac18\left(
    [\deltaR(z)+\deltaR(z')]^2\frac{d}{dR_0}\right. \\
  & \qquad\left.
    + \frac{[\deltaR(z)-\deltaR(z')]^2}{R_0}\right)\frac{d\Vo{z-z'}}{dR_0}.
\end{align*}
For future use, let us define the Fourier transform of $\Vo{z}$ as
\begin{equation}
  \Vq{q} = \int_0^L \!dz\,e^{-iqz}\Vo{z}.
\end{equation}
Note that $\text{Re}\Vq{q}>0$.

\subsection*{Inverse dielectric function $\epsilon^{-1}(z,z')$}\label{app:eps}

We first expand the dielectric function $\epsilon(z,z')$, Eq.\ (\ref{eq:dielectric1D}), as
\begin{multline}\label{eq:epsexpand}
  \epsilon(z,z') = \epso{z-z'} \\
    + \lambda\,\epsilon^{(1)}(z,z') + \lambda^2\,\epsilon^{(2)}(z,z'),
\end{multline}
where the zeroth-order term is
\begin{equation}
  \epso{z} = \delta(z)
    + \frac{g^{(0)}(\varepsilon_+)+g^{(0)}(\varepsilon_-)}{2}\Vo{z},
\end{equation}
the first-order term is
\begin{multline*}
  \epsilon^{(1)}(z,z^{\prime}) =\frac12\sum_{\alpha=\pm}\big(
    g^{(1)}(z,\varepsilon_\alpha)\Vo{z-z^{\prime}} \\
    + g^{(0)}(z,\varepsilon_\alpha)V^{(1)}(z,z^{\prime})
    \big),
\end{multline*}
and the second-order term is
\begin{multline*}
  \epsilon^{(2)}(z,z^{\prime}) =\frac12\sum_{\alpha=\pm}\Big(
    g^{(2)}(z,\varepsilon_\alpha)\Vo{z-z^{\prime}} \\
    + g^{(1)}(z,\varepsilon_\alpha)V^{(1)}(z,z^{\prime})
    + g^{(0)}(z,\varepsilon_\alpha)V^{(2)}(z,z^{\prime})
    \Big).
\end{multline*}
Let us define the Fourier transform of $\epso{z}$ as
\begin{equation}\begin{split}
  \epsq{q} &= \int_0^L \!dz\,e^{-iqz}\epso{z} \\
  &= 1+\frac{g^{(0)}(\varepsilon_+)+g^{(0)}(\varepsilon_-)}{2}\Vq{q}.
\end{split}\end{equation}
Note that $\text{Re}\left(\epsq{q}/\Vq{q}\right) >0$, since both $g^{(0)}(\varepsilon)>0$ and $\Vq{q} >0$.
Substituting Eq.\ (\ref{eq:epsexpand}) into the identity
\begin{equation*}
  \int\!dz^{\prime\prime}\,
    \epsilon^{-1}(z,z^{\prime\prime})\epsilon(z^{\prime\prime},z^{\prime})
  = \delta(z-z^{\prime}),
\end{equation*}
one can solve order by order for the inverse dielectric function $\epsilon^{-1}(z,z')$
\begin{multline}\label{eq:invepsexpand}
  \epsilon^{-1}(z,z') = \epsilon^{-1,(0)}(z-z') \\
    + \lambda\,\epsilon^{-1,(1)}(z,z')
    + \lambda^2\,\epsilon^{-1,(2)}(z,z'),
\end{multline}
where the zeroth-order term is
\begin{equation*}
  \epsilon^{-1,(0)}(z) = \frac{1}{L}\sum_q\frac{e^{iqz}}{\epsq{q}},
\end{equation*}
the first-order term is found to be
\begin{multline*}
  \epsilon^{-1,(1)} = -\int\!dz_1\!\int\!dz_2\,\epsilon^{-1,(0)}(z-z_1) \\
    \times\epsilon^{(1)}(z_1,z_2)\,\epsilon^{-1,(0)}(z_2-z'),
\end{multline*}
and the second-order term is
\begin{multline*}
  \epsilon^{-1,(2)}(z,z^{\prime}) = -\int\!dz_1\!\int\!dz_2\,
    \epsilon^{-1,(0)}(z_2-z^{\prime}) \\
    \times\bigg[\epsilon^{-1,(0)}(z-z_1)\,\epsilon^{(2)}(z_1,z_2) \\
    + \epsilon^{-1,(1)}(z-z_1)\,\epsilon^{(1)}(z_1,z_2) \bigg].
\end{multline*}

\subsection*{Screened Potential $\Vsc(z,z')$}\label{app:Vtilde}

Substituting the expansions (\ref{eq:Vexpand}) and (\ref{eq:invepsexpand}) of $V$ and $\epsilon^{-1}$ into Eq.\ (\ref{eq:screenedV}), one gets an expansion of the screened potential as
\begin{multline}\label{eq:Vtildeexpand}
  \Vsc(z,z') = \Vsc^{(0)}(z-z') \\
    + \lambda\,\Vsc^{(1)}(z,z')
    + \lambda^2\,\Vsc^{(2)}(z,z')
\end{multline}
where the zeroth-order term is
\begin{equation*}
  \Vsc^{(0)}(z) =
    \int\!dz_1\,\Vo{z-z_1}\epsilon^{-1,(0)}(z_1)
  = \frac{1}{L}\sum_q \frac{\Vq{q}}{\epsq{q}}
    e^{\displaystyle iqz},
\end{equation*}
the first-order term is
\begin{align*}
  \Vsc^{(1)}(z,z^{\prime})
    = \int\!dz_1\bigg[
      &\Vo{z-z_1}\epsilon^{-1,(1)}(z_1,z^{\prime}) \\
      &+ V^{(1)}(z-z_1)\epsilon^{-1,(0)}(z_1-z^{\prime})\bigg],
\end{align*}
while the second-order term is
\begin{align*}
  \Vsc^{(2)}(z,z^{\prime})
    = \int\!dz_1\bigg[&\Vo{z-z_1}\epsilon^{-1,(2)}(z_1,z^{\prime}) \\
      &+ V^{(1)}(z-z_1)\epsilon^{-1,(1)}(z_1,z^{\prime}) \\
      &+ V^{(2)}(z-z_1)\epsilon^{-1,(0)}(z_1-z^{\prime})
      \bigg].
\end{align*}

\subsection*{Grand canonical potential $\Omega[V,R(z),U(z)]$}\label{app:Omega}

Using Eqs.\ (\ref{eq:dosWeylexpand}), (\ref{eq:ddosexpand}), (\ref{eq:drhoexpand}), (\ref{eq:drhobarexpand}), and (\ref{eq:Vtildeexpand}) for
$g(z,E)$, $\delta\rho_0$, $\delta\bar{\rho}_0$, and $\Vsc$,
we are now ready to expand $\Omega$, starting by rewriting Eq.\ (\ref{eq:Omega1DexpandA}) as
\begin{multline}\label{eq:omrho0}
  \Omega \simeq \Omega_0[\{\mu_\pm\},R(z),U_0] - N_+U_0 \\
    - \Omega_1[\{\mu_\pm\},R(z),U_0] + \Omega_2[\{\mu_\pm\},R(z),U_0] \\
    + \Omega_3[\{\mu_\pm\},R(z),U_0],
\end{multline}
where
\begin{multline*}
 \Omega_1 = U_0\int\!dz\,\delta\bar{\rho}_0(z) \\
   \shoveleft{\quad\,\,
     = U_0L\Bigg[
     \frac{U_0}{2}\sum_{\alpha=\pm}g^{(0)}(\eps\alpha)
     + \lambda \bar\varrho^{(1)}_1b(0)} \\
     + \lambda^2 \sum_q\left[
       \bar\varrho^{(2)}_1
       + (\bar\varrho^{(2)}_2 - \bar\varrho^{(2)}_3)q^2
       \right] |b(q)|^2
     \Bigg],
\end{multline*}
\begin{multline*}
  \Omega_2 = \frac{(U_0)^2}{4}
    \sum_{\alpha=\pm}\int\!dz\,g(z,\varepsilon_\alpha) \\
   \shoveleft{\quad\,\,
     = \frac{(U_0)^2}{4}L
      \sum_{\alpha=\pm}\Bigg[g^{(0)}(\eps\alpha)
      + \lambda g^{(1)}_1 (\eps\alpha) b(0)} \\
      + \lambda^2 \sum_q\left[g^{(2)}_1(\eps\alpha)
        + [g^{(2)}_2(\eps\alpha) - g^{(2)}_3(\eps\alpha)]q^2\right] |b(q)|^2
      \Bigg],
\end{multline*}
and $\Omega_3 = \frac12\int dzdz^{\prime}\delta\bar{\rho}_0(z)
\Vsc(z,z^{\prime})\delta\bar{\rho}_0(z)$ can be written as
\begin{multline*}
  \Omega_3[\{\mu_\pm\},R(z),U_0] = \Omega_3^{(0)}[\{\mu_\pm\},R_0,U_0] \\
    + \lambda\,\Omega_{3}^{(1)}[\{\mu_\pm\},R_0,U_0]
    + \lambda^2\,\Omega_{3}^{(2)}[\{\mu_\pm\},R_0,U_0].
\end{multline*}
The zeroth order term in the expansion of $\Omega_3$ is
\begin{align*}
  \Omega^{(0)}_3 &= \frac12\int\!dzdz^{\prime}\,
    \delta\bar{\rho}^{(0)}_0(z)
    \Vsc^{(0)}(z-z^{\prime})\delta\bar{\rho}^{(0)}_0(z') \\
  &= \frac{L}{4}(U_0)^2\sum_{\alpha=\pm}g^{(0)}(\eps{\alpha}),
\end{align*}
the first-order term is
\begin{align*}
  \Omega^{(1)}_3 &= \frac12\int\!dzdz^{\prime}\,\bigg[
      \delta\bar{\rho}^{(0)}_0(z)
      \Vsc^{(1)}(z,z^{\prime})
      \delta\bar{\rho}^{(0)}_0(z^{\prime}) \\
    &\hspace{2.2cm}
      + 2\delta\bar{\rho}^{(1)}_0(z)
      \Vsc^{(0)}(z-z^{\prime})
      \delta\bar{\rho}^{(0)}_0(z^{\prime})
    \bigg] \\
    &= LU_0\bar\varrho_1^{(1)}b(0) + {\cal O}(L^0),
\end{align*}
and the second-order contribution is
\begin{alignat*}{2}
    \Omega^{(2)}_3 &= \frac12\int\!dzdz^{\prime}\,\Big[
      \delta\bar{\rho}^{(0)}_0(z)
      \Vsc^{(2)}(z,z^{\prime})
      \delta\bar{\rho}^{(0)}_0(z^{\prime}) \\
    &\hspace{2.2cm}
    + 2 \delta\bar{\rho}^{(0)}_0(z)
      \Vsc^{(1)}(z,z^{\prime})
      \delta\bar{\rho}^{(1)}_0(z^{\prime}) \\
    &\hspace{2.2cm}
    + 2 \delta\bar{\rho}^{(2)}_0(z)
      \Vsc^{(0)}(z-z^{\prime})
      \delta\bar{\rho}^{(0)}_0(z^{\prime}) \\
    &\hspace{2.2cm}
    + \delta\bar{\rho}^{(1)}_0(z)
      \Vsc^{(0)}(z-z^{\prime})
      \delta\bar{\rho}^{(1)}_0(z^{\prime})
    \Big] \displaybreak[1] \\
    &= LU_0\sum_q \Big[\varrho_1^{(2)}
       +(\varrho_2^{(2)}-\varrho_3^{(2)})q^2\Big] |b(q)|^2 \\
    &\quad+\frac{L}{2}\sum_q\frac{\Vq{q}}{\epsq{q}}
      \Big[\varrho_1^{(1)}-\varrho_2^{(1)}q^2\Big]^2 |b(q)|^2
    + {\cal O}(L^0).
\end{alignat*}
Adding up all the contributions in Eq.\ (\ref{eq:omrho0}), and
dropping contributions of order $L^0$, one gets Eqs.\ (\ref{eq:Omega1DexpandB}) and (\ref{eq:stabcoef}).

The above calculations also show that Eq.\ (\ref{eq:Omega1DexpandA}) can  be rewritten as
\begin{multline}
  \Omega = \Omega_0[\{\mu_\pm\},R(z),U_0] - N_+U_0 \\
  + \frac12\int\!dzdz^{\prime}\,
    \delta\rho_0^{(1)}(z)\Vsc^{(0)}(z-z^{\prime})
    \delta\rho_0^{(1)}(z^{\prime}) \\
  + {\cal O}(L^0).
\end{multline}

\bibliography{zhang05}

\begin{thebibliography}{53}
\expandafter\ifx\csname natexlab\endcsname\relax\def\natexlab#1{#1}\fi
\expandafter\ifx\csname bibnamefont\endcsname\relax
  \def\bibnamefont#1{#1}\fi
\expandafter\ifx\csname bibfnamefont\endcsname\relax
  \def\bibfnamefont#1{#1}\fi
\expandafter\ifx\csname citenamefont\endcsname\relax
  \def\citenamefont#1{#1}\fi
\expandafter\ifx\csname url\endcsname\relax
  \def\url#1{\texttt{#1}}\fi
\expandafter\ifx\csname urlprefix\endcsname\relax\def\urlprefix{URL }\fi
\providecommand{\bibinfo}[2]{#2}
\providecommand{\eprint}[2][]{\url{#2}}

\bibitem[{\citenamefont{Agra{\"\i}t et~al.}(2003)\citenamefont{Agra{\"\i}t,
  Levy~Yeyati, and van Ruitenbeek}}]{Agrait03}
\bibinfo{author}{\bibfnamefont{N.}~\bibnamefont{Agra{\"\i}t}},
  \bibinfo{author}{\bibfnamefont{A.}~\bibnamefont{Levy~Yeyati}},
  \bibnamefont{and} \bibinfo{author}{\bibfnamefont{J.~M.} \bibnamefont{van
  Ruitenbeek}}, \bibinfo{journal}{Phys. Rep.} \textbf{\bibinfo{volume}{377}},
  \bibinfo{pages}{81} (\bibinfo{year}{2003}).

\bibitem[{\citenamefont{Yasuda and Sakai}(1997)}]{Yasuda97}
\bibinfo{author}{\bibfnamefont{H.}~\bibnamefont{Yasuda}} \bibnamefont{and}
  \bibinfo{author}{\bibfnamefont{A.}~\bibnamefont{Sakai}},
  \bibinfo{journal}{Phys. Rev. B} \textbf{\bibinfo{volume}{56}},
  \bibinfo{pages}{1069} (\bibinfo{year}{1997}).

\bibitem[{\citenamefont{Itakura et~al.}(1999)\citenamefont{Itakura, Yuki,
  Kurokawa, Yasuda, and Sakai}}]{Itakura99}
\bibinfo{author}{\bibfnamefont{K.}~\bibnamefont{Itakura}},
  \bibinfo{author}{\bibfnamefont{K.}~\bibnamefont{Yuki}},
  \bibinfo{author}{\bibfnamefont{S.}~\bibnamefont{Kurokawa}},
  \bibinfo{author}{\bibfnamefont{H.}~\bibnamefont{Yasuda}}, \bibnamefont{and}
  \bibinfo{author}{\bibfnamefont{A.}~\bibnamefont{Sakai}},
  \bibinfo{journal}{Phys. Rev. B} \textbf{\bibinfo{volume}{60}},
  \bibinfo{pages}{11163} (\bibinfo{year}{1999}).

\bibitem[{\citenamefont{Untiedt et~al.}(2000)\citenamefont{Untiedt,
  Rubio~Bollinger, Vieira, and Agra{\"\i}t}}]{Untiedt00}
\bibinfo{author}{\bibfnamefont{C.}~\bibnamefont{Untiedt}},
  \bibinfo{author}{\bibfnamefont{G.}~\bibnamefont{Rubio~Bollinger}},
  \bibinfo{author}{\bibfnamefont{S.}~\bibnamefont{Vieira}}, \bibnamefont{and}
  \bibinfo{author}{\bibfnamefont{N.}~\bibnamefont{Agra{\"\i}t}},
  \bibinfo{journal}{Phys. Rev. B} \textbf{\bibinfo{volume}{62}},
  \bibinfo{pages}{9962} (\bibinfo{year}{2000}).

\bibitem[{\citenamefont{Hansen et~al.}(2000)\citenamefont{Hansen, Nielsen,
  Brandbyge, L{\ae}gsgaard, Stensgaard, and Besenbacher}}]{Hansen00}
\bibinfo{author}{\bibfnamefont{K.}~\bibnamefont{Hansen}},
  \bibinfo{author}{\bibfnamefont{S.~K.} \bibnamefont{Nielsen}},
  \bibinfo{author}{\bibfnamefont{M.}~\bibnamefont{Brandbyge}},
  \bibinfo{author}{\bibfnamefont{E.}~\bibnamefont{L{\ae}gsgaard}},
  \bibinfo{author}{\bibfnamefont{I.}~\bibnamefont{Stensgaard}},
  \bibnamefont{and}
  \bibinfo{author}{\bibfnamefont{F.}~\bibnamefont{Besenbacher}},
  \bibinfo{journal}{Appl. Phys. Lett.} \textbf{\bibinfo{volume}{77}},
  \bibinfo{pages}{708} (\bibinfo{year}{2000}).

\bibitem[{\citenamefont{Yuki et~al.}(2001)\citenamefont{Yuki, Enomoto, and
  Sakai}}]{Yuki01}
\bibinfo{author}{\bibfnamefont{K.}~\bibnamefont{Yuki}},
  \bibinfo{author}{\bibfnamefont{A.}~\bibnamefont{Enomoto}}, \bibnamefont{and}
  \bibinfo{author}{\bibfnamefont{A.}~\bibnamefont{Sakai}},
  \bibinfo{journal}{Appl. Surf. Sci.} \textbf{\bibinfo{volume}{169--170}},
  \bibinfo{pages}{489} (\bibinfo{year}{2001}).

\bibitem[{\citenamefont{Mehrez et~al.}(2002)\citenamefont{Mehrez, Wlasenko,
  Larade, Taylor, Grutter, and Guo}}]{Mehrez02}
\bibinfo{author}{\bibfnamefont{H.}~\bibnamefont{Mehrez}},
  \bibinfo{author}{\bibfnamefont{A.}~\bibnamefont{Wlasenko}},
  \bibinfo{author}{\bibfnamefont{B.}~\bibnamefont{Larade}},
  \bibinfo{author}{\bibfnamefont{J.}~\bibnamefont{Taylor}},
  \bibinfo{author}{\bibfnamefont{P.}~\bibnamefont{Grutter}}, \bibnamefont{and}
  \bibinfo{author}{\bibfnamefont{H.}~\bibnamefont{Guo}},
  \bibinfo{journal}{Phys. Rev. B} \textbf{\bibinfo{volume}{65}},
  \bibinfo{pages}{195419} (\bibinfo{year}{2002}).

\bibitem[{\citenamefont{Agra{\"\i}t et~al.}(2002)\citenamefont{Agra{\"\i}t,
  Untiedt, Rubio-Bollinger, and Vieira}}]{Agrait02}
\bibinfo{author}{\bibfnamefont{N.}~\bibnamefont{Agra{\"\i}t}},
  \bibinfo{author}{\bibfnamefont{C.}~\bibnamefont{Untiedt}},
  \bibinfo{author}{\bibfnamefont{G.}~\bibnamefont{Rubio-Bollinger}},
  \bibnamefont{and} \bibinfo{author}{\bibfnamefont{S.}~\bibnamefont{Vieira}},
  \bibinfo{journal}{Phys. Rev. Lett.} \textbf{\bibinfo{volume}{88}},
  \bibinfo{pages}{216803} (\bibinfo{year}{2002}).

\bibitem[{\citenamefont{Pascual et~al.}(1997)\citenamefont{Pascual, Torres, and
  S\'aenz}}]{Pascual97}
\bibinfo{author}{\bibfnamefont{J.~I.} \bibnamefont{Pascual}},
  \bibinfo{author}{\bibfnamefont{J.~A.} \bibnamefont{Torres}},
  \bibnamefont{and} \bibinfo{author}{\bibfnamefont{J.~J.}
  \bibnamefont{S\'aenz}}, \bibinfo{journal}{Phys. Rev. B}
  \textbf{\bibinfo{volume}{55}}, \bibinfo{pages}{16029} (\bibinfo{year}{1997}).

\bibitem[{\citenamefont{Bogacheck et~al.}(1997)\citenamefont{Bogacheck,
  Scherbakov, and Landman}}]{Bogachek97}
\bibinfo{author}{\bibfnamefont{E.~N.} \bibnamefont{Bogacheck}},
  \bibinfo{author}{\bibfnamefont{A.~G.} \bibnamefont{Scherbakov}},
  \bibnamefont{and} \bibinfo{author}{\bibfnamefont{U.}~\bibnamefont{Landman}},
  \bibinfo{journal}{Phys. Rev. B} \textbf{\bibinfo{volume}{56}},
  \bibinfo{pages}{14917} (\bibinfo{year}{1997}).

\bibitem[{\citenamefont{Zagoskin}(1998)}]{Zagoskin98}
\bibinfo{author}{\bibfnamefont{A.~M.} \bibnamefont{Zagoskin}},
  \bibinfo{journal}{Phys. Rev. B} \textbf{\bibinfo{volume}{58}},
  \bibinfo{pages}{15827} (\bibinfo{year}{1998}).

\bibitem[{\citenamefont{Christen and B\"uttiker}(1996)}]{Christen96}
\bibinfo{author}{\bibfnamefont{T.}~\bibnamefont{Christen}} \bibnamefont{and}
  \bibinfo{author}{\bibfnamefont{M.}~\bibnamefont{B\"uttiker}},
  \bibinfo{journal}{Europhys. Lett.} \textbf{\bibinfo{volume}{35}},
  \bibinfo{pages}{523} (\bibinfo{year}{1996}).

\bibitem[{\citenamefont{Todorov et~al.}(2000)\citenamefont{Todorov, Hoekstra,
  and Sutton}}]{Todorov00}
\bibinfo{author}{\bibfnamefont{T.~N.} \bibnamefont{Todorov}},
  \bibinfo{author}{\bibfnamefont{J.}~\bibnamefont{Hoekstra}}, \bibnamefont{and}
  \bibinfo{author}{\bibfnamefont{A.~P.} \bibnamefont{Sutton}},
  \bibinfo{journal}{Phil. Mag. B} \textbf{\bibinfo{volume}{80}},
  \bibinfo{pages}{421} (\bibinfo{year}{2000}).

\bibitem[{\citenamefont{Di~Ventra and Lang}(2002)}]{Ventra01}
\bibinfo{author}{\bibfnamefont{M.}~\bibnamefont{Di~Ventra}} \bibnamefont{and}
  \bibinfo{author}{\bibfnamefont{N.~D.} \bibnamefont{Lang}},
  \bibinfo{journal}{Phys. Rev. B} \textbf{\bibinfo{volume}{65}},
  \bibinfo{pages}{045402} (\bibinfo{year}{2002}).

\bibitem[{\citenamefont{Brandbyge et~al.}(2002)\citenamefont{Brandbyge, Mozos,
  Ordej\'on, Taylor, and Stokbro}}]{Brandbyge02}
\bibinfo{author}{\bibfnamefont{M.}~\bibnamefont{Brandbyge}},
  \bibinfo{author}{\bibfnamefont{J.-L.} \bibnamefont{Mozos}},
  \bibinfo{author}{\bibfnamefont{P.}~\bibnamefont{Ordej\'on}},
  \bibinfo{author}{\bibfnamefont{J.}~\bibnamefont{Taylor}}, \bibnamefont{and}
  \bibinfo{author}{\bibfnamefont{K.}~\bibnamefont{Stokbro}},
  \bibinfo{journal}{Phys. Rev. B} \textbf{\bibinfo{volume}{65}},
  \bibinfo{pages}{165401} (\bibinfo{year}{2002}).

\bibitem[{\citenamefont{Mozos et~al.}(2002)\citenamefont{Mozos, Ordej\'on,
  Brandbyge, Taylor, and Stokbro}}]{Mozos02}
\bibinfo{author}{\bibfnamefont{J.~L.} \bibnamefont{Mozos}},
  \bibinfo{author}{\bibfnamefont{P.}~\bibnamefont{Ordej\'on}},
  \bibinfo{author}{\bibfnamefont{M.}~\bibnamefont{Brandbyge}},
  \bibinfo{author}{\bibfnamefont{J.}~\bibnamefont{Taylor}}, \bibnamefont{and}
  \bibinfo{author}{\bibfnamefont{K.}~\bibnamefont{Stokbro}},
  \bibinfo{journal}{Nanotechnology} \textbf{\bibinfo{volume}{13}},
  \bibinfo{pages}{346} (\bibinfo{year}{2002}).

\bibitem[{\citenamefont{Urban et~al.}(2004{\natexlab{a}})\citenamefont{Urban,
  B\"urki, Yanson, Yanson, Stafford, van Ruitenbeek, and Grabert}}]{Urban04a}
\bibinfo{author}{\bibfnamefont{D.~F.} \bibnamefont{Urban}},
  \bibinfo{author}{\bibfnamefont{J.}~\bibnamefont{B\"urki}},
  \bibinfo{author}{\bibfnamefont{A.~I.} \bibnamefont{Yanson}},
  \bibinfo{author}{\bibfnamefont{I.~K.} \bibnamefont{Yanson}},
  \bibinfo{author}{\bibfnamefont{C.~A.} \bibnamefont{Stafford}},
  \bibinfo{author}{\bibfnamefont{J.~M.} \bibnamefont{van Ruitenbeek}},
  \bibnamefont{and} \bibinfo{author}{\bibfnamefont{H.}~\bibnamefont{Grabert}},
  \bibinfo{journal}{Solid State Comm.} \textbf{\bibinfo{volume}{131}},
  \bibinfo{pages}{609} (\bibinfo{year}{2004}{\natexlab{a}}).

\bibitem[{\citenamefont{Stafford et~al.}(1997)\citenamefont{Stafford,
  Baeriswyl, and B\"urki}}]{Stafford97a}
\bibinfo{author}{\bibfnamefont{C.~A.} \bibnamefont{Stafford}},
  \bibinfo{author}{\bibfnamefont{D.}~\bibnamefont{Baeriswyl}},
  \bibnamefont{and} \bibinfo{author}{\bibfnamefont{J.}~\bibnamefont{B\"urki}},
  \bibinfo{journal}{Phys. Rev. Lett.} \textbf{\bibinfo{volume}{79}},
  \bibinfo{pages}{2863} (\bibinfo{year}{1997}).

\bibitem[{\citenamefont{Kassubek et~al.}(1999)\citenamefont{Kassubek, Stafford,
  and Grabert}}]{Kassubek99}
\bibinfo{author}{\bibfnamefont{F.}~\bibnamefont{Kassubek}},
  \bibinfo{author}{\bibfnamefont{C.~A.} \bibnamefont{Stafford}},
  \bibnamefont{and} \bibinfo{author}{\bibfnamefont{H.}~\bibnamefont{Grabert}},
  \bibinfo{journal}{Phys. Rev. B} \textbf{\bibinfo{volume}{59}},
  \bibinfo{pages}{7560} (\bibinfo{year}{1999}).

\bibitem[{\citenamefont{Stafford et~al.}(1999)\citenamefont{Stafford, Kassubek,
  B\"urki, and Grabert}}]{Stafford99}
\bibinfo{author}{\bibfnamefont{C.~A.} \bibnamefont{Stafford}},
  \bibinfo{author}{\bibfnamefont{F.}~\bibnamefont{Kassubek}},
  \bibinfo{author}{\bibfnamefont{J.}~\bibnamefont{B\"urki}}, \bibnamefont{and}
  \bibinfo{author}{\bibfnamefont{H.}~\bibnamefont{Grabert}},
  \bibinfo{journal}{Phys. Rev. Lett.} \textbf{\bibinfo{volume}{83}},
  \bibinfo{pages}{4836} (\bibinfo{year}{1999}).

\bibitem[{\citenamefont{Kassubek et~al.}(2001)\citenamefont{Kassubek, Stafford,
  Grabert, and Goldstein}}]{Kassubek01}
\bibinfo{author}{\bibfnamefont{F.}~\bibnamefont{Kassubek}},
  \bibinfo{author}{\bibfnamefont{C.~A.} \bibnamefont{Stafford}},
  \bibinfo{author}{\bibfnamefont{H.}~\bibnamefont{Grabert}}, \bibnamefont{and}
  \bibinfo{author}{\bibfnamefont{R.~E.} \bibnamefont{Goldstein}},
  \bibinfo{journal}{Nonlinearity} \textbf{\bibinfo{volume}{14}},
  \bibinfo{pages}{167} (\bibinfo{year}{2001}).

\bibitem[{\citenamefont{Zhang et~al.}(2003)\citenamefont{Zhang, Kassubek, and
  Stafford}}]{Zhang03}
\bibinfo{author}{\bibfnamefont{C.-H.} \bibnamefont{Zhang}},
  \bibinfo{author}{\bibfnamefont{F.}~\bibnamefont{Kassubek}}, \bibnamefont{and}
  \bibinfo{author}{\bibfnamefont{C.~A.} \bibnamefont{Stafford}},
  \bibinfo{journal}{Phys. Rev. B} \textbf{\bibinfo{volume}{68}},
  \bibinfo{pages}{165414} (\bibinfo{year}{2003}).

\bibitem[{\citenamefont{B\"urki et~al.}(2003)\citenamefont{B\"urki, Goldstein,
  and Stafford}}]{Burki03}
\bibinfo{author}{\bibfnamefont{J.}~\bibnamefont{B\"urki}},
  \bibinfo{author}{\bibfnamefont{R.~E.} \bibnamefont{Goldstein}},
  \bibnamefont{and} \bibinfo{author}{\bibfnamefont{C.~A.}
  \bibnamefont{Stafford}}, \bibinfo{journal}{Phys. Rev. Lett.}
  \textbf{\bibinfo{volume}{91}}, \bibinfo{pages}{254501}
  (\bibinfo{year}{2003}).

\bibitem[{\citenamefont{Urban et~al.}(2004{\natexlab{b}})\citenamefont{Urban,
  B\"urki, Zhang, Stafford, and Grabert}}]{Urban04b}
\bibinfo{author}{\bibfnamefont{D.~F.} \bibnamefont{Urban}},
  \bibinfo{author}{\bibfnamefont{J.}~\bibnamefont{B\"urki}},
  \bibinfo{author}{\bibfnamefont{C.-H.} \bibnamefont{Zhang}},
  \bibinfo{author}{\bibfnamefont{C.~A.} \bibnamefont{Stafford}},
  \bibnamefont{and} \bibinfo{author}{\bibfnamefont{H.}~\bibnamefont{Grabert}},
  \bibinfo{journal}{Phys. Rev. Lett.} \textbf{\bibinfo{volume}{93}},
  \bibinfo{pages}{186403} (\bibinfo{year}{2004}{\natexlab{b}}).

\bibitem[{\citenamefont{Stafford et~al.}(2000)\citenamefont{Stafford, Kassubek,
  B\"urki, Grabert, and Baeriswyl}}]{Stafford00}
\bibinfo{author}{\bibfnamefont{C.~A.} \bibnamefont{Stafford}},
  \bibinfo{author}{\bibfnamefont{F.}~\bibnamefont{Kassubek}},
  \bibinfo{author}{\bibfnamefont{J.}~\bibnamefont{B\"urki}},
  \bibinfo{author}{\bibfnamefont{H.}~\bibnamefont{Grabert}}, \bibnamefont{and}
  \bibinfo{author}{\bibfnamefont{D.}~\bibnamefont{Baeriswyl}}, in
  \emph{\bibinfo{booktitle}{Quantum Physics at the Mesoscopic Scale}}, edited
  by \bibinfo{editor}{\bibfnamefont{D.~C.} \bibnamefont{Glattli}},
  \bibinfo{editor}{\bibfnamefont{M.}~\bibnamefont{Sanquer}}, \bibnamefont{and}
  \bibinfo{editor}{\bibfnamefont{J.}~\bibnamefont{Tran Thanh~Van}}
  (\bibinfo{publisher}{EDP Sciences}, \bibinfo{address}{Les Ulis, France},
  \bibinfo{year}{2000}), pp. \bibinfo{pages}{445--449}.

\bibitem[{\citenamefont{Stafford et~al.}(2001)\citenamefont{Stafford, Kassubek,
  and Grabert}}]{Stafford01}
\bibinfo{author}{\bibfnamefont{C.~A.} \bibnamefont{Stafford}},
  \bibinfo{author}{\bibfnamefont{F.}~\bibnamefont{Kassubek}}, \bibnamefont{and}
  \bibinfo{author}{\bibfnamefont{H.}~\bibnamefont{Grabert}}, in
  \emph{\bibinfo{booktitle}{Advances in Solid State Physics}}, edited by
  \bibinfo{editor}{\bibfnamefont{B.}~\bibnamefont{Kramer}}
  (\bibinfo{publisher}{Springer-Verlag}, \bibinfo{address}{Berlin Heidelberg},
  \bibinfo{year}{2001}), pp. \bibinfo{pages}{497--511}.

\bibitem[{\citenamefont{Christen}(1996)}]{Christen97}
\bibinfo{author}{\bibfnamefont{T.}~\bibnamefont{Christen}},
  \bibinfo{journal}{Phys. Rev. B} \textbf{\bibinfo{volume}{55}},
  \bibinfo{pages}{7606} (\bibinfo{year}{1996}).

\bibitem[{\citenamefont{B\"uttiker}(1993)}]{Buttiker93}
\bibinfo{author}{\bibfnamefont{M.}~\bibnamefont{B\"uttiker}},
  \bibinfo{journal}{J. Phys. Condens. Matter} \textbf{\bibinfo{volume}{5}},
  \bibinfo{pages}{9361} (\bibinfo{year}{1993}).

\bibitem[{\citenamefont{Gasparian et~al.}(1996)\citenamefont{Gasparian,
  Christen, and B\"uttiker}}]{Gasparian96}
\bibinfo{author}{\bibfnamefont{V.}~\bibnamefont{Gasparian}},
  \bibinfo{author}{\bibfnamefont{T.}~\bibnamefont{Christen}}, \bibnamefont{and}
  \bibinfo{author}{\bibfnamefont{M.}~\bibnamefont{B\"uttiker}},
  \bibinfo{journal}{Phys. Rev. A} \textbf{\bibinfo{volume}{54}},
  \bibinfo{pages}{4022} (\bibinfo{year}{1996}).

\bibitem[{not({\natexlab{a}})}]{note_hardwall}
\bibinfo{note}{Other choices for the transverse confinement
  potential\cite{Garcia_Martin96,Yannouleas98,Puska01} lead to qualitatively
  similar results.}

\bibitem[{\citenamefont{Garc{\'\i}a-Mart{\'\i}n
  et~al.}(1996)\citenamefont{Garc{\'\i}a-Mart{\'\i}n, Torres, and
  S\'aenz}}]{Garcia_Martin96}
\bibinfo{author}{\bibfnamefont{A.}~\bibnamefont{Garc{\'\i}a-Mart{\'\i}n}},
  \bibinfo{author}{\bibfnamefont{J.~A.} \bibnamefont{Torres}},
  \bibnamefont{and} \bibinfo{author}{\bibfnamefont{J.~J.}
  \bibnamefont{S\'aenz}}, \bibinfo{journal}{Phys. Rev. B}
  \textbf{\bibinfo{volume}{54}}, \bibinfo{pages}{13448} (\bibinfo{year}{1996}).

\bibitem[{\citenamefont{Yannouleas et~al.}(1998)\citenamefont{Yannouleas,
  Bogachek, and Landman}}]{Yannouleas98}
\bibinfo{author}{\bibfnamefont{C.}~\bibnamefont{Yannouleas}},
  \bibinfo{author}{\bibfnamefont{E.~N.} \bibnamefont{Bogachek}},
  \bibnamefont{and} \bibinfo{author}{\bibfnamefont{U.}~\bibnamefont{Landman}},
  \bibinfo{journal}{Phys. Rev. B} \textbf{\bibinfo{volume}{57}},
  \bibinfo{pages}{4872} (\bibinfo{year}{1998}).

\bibitem[{\citenamefont{Puska et~al.}(2001)\citenamefont{Puska, Ogando, and
  Zabala}}]{Puska01}
\bibinfo{author}{\bibfnamefont{M.~J.} \bibnamefont{Puska}},
  \bibinfo{author}{\bibfnamefont{E.}~\bibnamefont{Ogando}}, \bibnamefont{and}
  \bibinfo{author}{\bibfnamefont{N.}~\bibnamefont{Zabala}},
  \bibinfo{journal}{Phys. Rev. B} \textbf{\bibinfo{volume}{64}},
  \bibinfo{pages}{033401} (\bibinfo{year}{2001}).

\bibitem[{\citenamefont{Lang}(1973)}]{Lang73}
\bibinfo{author}{\bibfnamefont{N.~D.} \bibnamefont{Lang}},
  \bibinfo{journal}{Solid State Physics} \textbf{\bibinfo{volume}{28}},
  \bibinfo{pages}{225} (\bibinfo{year}{1973}).

\bibitem[{\citenamefont{Torres et~al.}(1994)\citenamefont{Torres, Pascual, and
  S\'aenz}}]{Torres94}
\bibinfo{author}{\bibfnamefont{J.~A.} \bibnamefont{Torres}},
  \bibinfo{author}{\bibfnamefont{J.~I.} \bibnamefont{Pascual}},
  \bibnamefont{and} \bibinfo{author}{\bibfnamefont{J.~J.}
  \bibnamefont{S\'aenz}}, \bibinfo{journal}{Phys. Rev. B}
  \textbf{\bibinfo{volume}{49}}, \bibinfo{pages}{16581} (\bibinfo{year}{1994}).

\bibitem[{\citenamefont{B\"urki et~al.}(1999)\citenamefont{B\"urki, Stafford,
  Zotos, and Baeriswyl}}]{Burki99}
\bibinfo{author}{\bibfnamefont{J.}~\bibnamefont{B\"urki}},
  \bibinfo{author}{\bibfnamefont{C.~A.} \bibnamefont{Stafford}},
  \bibinfo{author}{\bibfnamefont{X.}~\bibnamefont{Zotos}}, \bibnamefont{and}
  \bibinfo{author}{\bibfnamefont{D.}~\bibnamefont{Baeriswyl}},
  \bibinfo{journal}{Phys. Rev. B} \textbf{\bibinfo{volume}{60}},
  \bibinfo{pages}{5000} (\bibinfo{year}{1999}).

\bibitem[{\citenamefont{B\"urki and Stafford}(1999)}]{Burki99b}
\bibinfo{author}{\bibfnamefont{J.}~\bibnamefont{B\"urki}} \bibnamefont{and}
  \bibinfo{author}{\bibfnamefont{C.~A.} \bibnamefont{Stafford}},
  \bibinfo{journal}{Phys. Rev. Lett.} \textbf{\bibinfo{volume}{83}},
  \bibinfo{pages}{3342} (\bibinfo{year}{1999}).

\bibitem[{\citenamefont{Brack and Bhaduri}(1997)}]{Brack97}
\bibinfo{author}{\bibfnamefont{M.}~\bibnamefont{Brack}} \bibnamefont{and}
  \bibinfo{author}{\bibfnamefont{R.~K.} \bibnamefont{Bhaduri}},
  \emph{\bibinfo{title}{Semiclassical Physics}}
  (\bibinfo{publisher}{Addison-Wesley}, \bibinfo{address}{Reading, MA},
  \bibinfo{year}{1997}).

\bibitem[{\citenamefont{Balian and Bloch}(1972)}]{Balian72}
\bibinfo{author}{\bibfnamefont{R.}~\bibnamefont{Balian}} \bibnamefont{and}
  \bibinfo{author}{\bibfnamefont{C.}~\bibnamefont{Bloch}},
  \bibinfo{journal}{Ann. Phys. (N. Y.)} \textbf{\bibinfo{volume}{69}},
  \bibinfo{pages}{76} (\bibinfo{year}{1972}).

\bibitem[{\citenamefont{Urban and Grabert}(2003)}]{Urban03}
\bibinfo{author}{\bibfnamefont{D.~F.} \bibnamefont{Urban}} \bibnamefont{and}
  \bibinfo{author}{\bibfnamefont{H.}~\bibnamefont{Grabert}},
  \bibinfo{journal}{Phys. Rev. Lett.} \textbf{\bibinfo{volume}{91}},
  \bibinfo{pages}{256803} (\bibinfo{year}{2003}).

\bibitem[{\citenamefont{Blom et~al.}(1998)\citenamefont{Blom, Olin,
  Costa-Kr\"amer, Garc{\'\i}a, Jonson, A., and Shekhter}}]{Blom98}
\bibinfo{author}{\bibfnamefont{S.}~\bibnamefont{Blom}},
  \bibinfo{author}{\bibfnamefont{H.}~\bibnamefont{Olin}},
  \bibinfo{author}{\bibfnamefont{J.~L.} \bibnamefont{Costa-Kr\"amer}},
  \bibinfo{author}{\bibfnamefont{N.}~\bibnamefont{Garc{\'\i}a}},
  \bibinfo{author}{\bibfnamefont{M.}~\bibnamefont{Jonson}},
  \bibinfo{author}{\bibfnamefont{S.~P.} \bibnamefont{A.}}, \bibnamefont{and}
  \bibinfo{author}{\bibfnamefont{R.~I.} \bibnamefont{Shekhter}},
  \bibinfo{journal}{Phys. Rev. B} \textbf{\bibinfo{volume}{57}},
  \bibinfo{pages}{8830} (\bibinfo{year}{1998}).

\bibitem[{\citenamefont{H\"oppler and Zwerger}(1999)}]{Hoeppler99}
\bibinfo{author}{\bibfnamefont{C.}~\bibnamefont{H\"oppler}} \bibnamefont{and}
  \bibinfo{author}{\bibfnamefont{W.}~\bibnamefont{Zwerger}},
  \bibinfo{journal}{Phys. Rev. B} \textbf{\bibinfo{volume}{59}},
  \bibinfo{pages}{R7849} (\bibinfo{year}{1999}).

\bibitem[{\citenamefont{van Ruitenbeek et~al.}(1997)\citenamefont{van
  Ruitenbeek, Devoret, Esteve, and Urbina}}]{Ruitenbeek97}
\bibinfo{author}{\bibfnamefont{J.~M.} \bibnamefont{van Ruitenbeek}},
  \bibinfo{author}{\bibfnamefont{M.~H.} \bibnamefont{Devoret}},
  \bibinfo{author}{\bibfnamefont{D.}~\bibnamefont{Esteve}}, \bibnamefont{and}
  \bibinfo{author}{\bibfnamefont{C.}~\bibnamefont{Urbina}},
  \bibinfo{journal}{Phys. Rev. B} \textbf{\bibinfo{volume}{56}},
  \bibinfo{pages}{12566} (\bibinfo{year}{1997}).

\bibitem[{\citenamefont{Rubio et~al.}(1996)\citenamefont{Rubio, Agra{\"\i}t,
  and Vieira}}]{Agrait96}
\bibinfo{author}{\bibfnamefont{C.}~\bibnamefont{Rubio}},
  \bibinfo{author}{\bibfnamefont{N.}~\bibnamefont{Agra{\"\i}t}},
  \bibnamefont{and} \bibinfo{author}{\bibfnamefont{S.}~\bibnamefont{Vieira}},
  \bibinfo{journal}{Phys. Rev. Lett.} \textbf{\bibinfo{volume}{76}},
  \bibinfo{pages}{2302} (\bibinfo{year}{1996}).

\bibitem[{\citenamefont{Stalder and D\"urig}(1996)}]{Stalder96}
\bibinfo{author}{\bibfnamefont{A.}~\bibnamefont{Stalder}} \bibnamefont{and}
  \bibinfo{author}{\bibfnamefont{U.}~\bibnamefont{D\"urig}},
  \bibinfo{journal}{Appl. Phys. Lett.} \textbf{\bibinfo{volume}{68}},
  \bibinfo{pages}{637} (\bibinfo{year}{1996}).

\bibitem[{\citenamefont{Rubio-Bollinger
  et~al.}(2004)\citenamefont{Rubio-Bollinger, Joyez, and Agrait}}]{Rubio04}
\bibinfo{author}{\bibfnamefont{G.}~\bibnamefont{Rubio-Bollinger}},
  \bibinfo{author}{\bibfnamefont{P.}~\bibnamefont{Joyez}}, \bibnamefont{and}
  \bibinfo{author}{\bibfnamefont{N.}~\bibnamefont{Agrait}},
  \bibinfo{journal}{Phys. Rev. Lett.} \textbf{\bibinfo{volume}{93}},
  \bibinfo{pages}{116803} (\bibinfo{year}{2004}).

\bibitem[{\citenamefont{Zhang}()}]{Zhang05}
\bibinfo{author}{\bibfnamefont{C.-H.} \bibnamefont{Zhang}},
  \bibinfo{howpublished}{unpublished}.

\bibitem[{\citenamefont{Strutinsky}(1968)}]{Strut68}
\bibinfo{author}{\bibfnamefont{V.~M.} \bibnamefont{Strutinsky}},
  \bibinfo{journal}{Nucl. Phys. A} \textbf{\bibinfo{volume}{122}},
  \bibinfo{pages}{1} (\bibinfo{year}{1968}).

\bibitem[{\citenamefont{Ullmo et~al.}(2001)\citenamefont{Ullmo, Nagano,
  Tomsovic, and Baranger}}]{Ullmo01}
\bibinfo{author}{\bibfnamefont{D.}~\bibnamefont{Ullmo}},
  \bibinfo{author}{\bibfnamefont{T.}~\bibnamefont{Nagano}},
  \bibinfo{author}{\bibfnamefont{S.}~\bibnamefont{Tomsovic}}, \bibnamefont{and}
  \bibinfo{author}{\bibfnamefont{H.~U.} \bibnamefont{Baranger}},
  \bibinfo{journal}{Phys. Rev. B} \textbf{\bibinfo{volume}{63}},
  \bibinfo{pages}{125339} (\bibinfo{year}{2001}).

\bibitem[{not({\natexlab{b}})}]{note_Peierls}
\bibinfo{note}{A short-wavelength Peierls instability also arises in a fully
  quantum-mechanical treatment,\cite{Urban03} but only at very low
  temperatures.}

\bibitem[{\citenamefont{Ralls et~al.}(1989)\citenamefont{Ralls, Ralph, and
  Buhrman}}]{Ralls89}
\bibinfo{author}{\bibfnamefont{K.~S.} \bibnamefont{Ralls}},
  \bibinfo{author}{\bibfnamefont{D.~C.} \bibnamefont{Ralph}}, \bibnamefont{and}
  \bibinfo{author}{\bibfnamefont{R.~A.} \bibnamefont{Buhrman}},
  \bibinfo{journal}{Phys. Rev. B} \textbf{\bibinfo{volume}{40}},
  \bibinfo{pages}{11561} (\bibinfo{year}{1989}).

\bibitem[{\citenamefont{Holweg et~al.}(1992)\citenamefont{Holweg, Caro,
  Verbruggen, and Radelaar}}]{Holweg92}
\bibinfo{author}{\bibfnamefont{P.~A.~M.} \bibnamefont{Holweg}},
  \bibinfo{author}{\bibfnamefont{J.}~\bibnamefont{Caro}},
  \bibinfo{author}{\bibfnamefont{A.~H.} \bibnamefont{Verbruggen}},
  \bibnamefont{and} \bibinfo{author}{\bibfnamefont{S.}~\bibnamefont{Radelaar}},
  \bibinfo{journal}{Phys. Rev. B} \textbf{\bibinfo{volume}{45}},
  \bibinfo{pages}{9311} (\bibinfo{year}{1992}).

\bibitem[{\citenamefont{B\"urki et~al.}(2004)\citenamefont{B\"urki, Stafford,
  and Stein}}]{Burki04}
\bibinfo{author}{\bibfnamefont{J.}~\bibnamefont{B\"urki}},
  \bibinfo{author}{\bibfnamefont{C.~A.} \bibnamefont{Stafford}},
  \bibnamefont{and} \bibinfo{author}{\bibfnamefont{D.~L.} \bibnamefont{Stein}},
  in \emph{\bibinfo{booktitle}{Noise in Complex Systems and Stochastic Dynamics
  II}}, edited by \bibinfo{editor}{\bibfnamefont{Z.}~\bibnamefont{Gingl}}
  (\bibinfo{publisher}{SPIE Publishing}, \bibinfo{address}{Bellingham, WA},
  \bibinfo{year}{2004}), vol. \bibinfo{volume}{5471}, pp.
  \bibinfo{pages}{367--379}.

\end{thebibliography}

\end{document}